\documentclass[aip,11pt,floatfix]{revtex4-2}
\usepackage[percent]{overpic}
\usepackage{graphicx}
\usepackage{dcolumn}
\usepackage{bm}
\usepackage{xcolor, soul}
\usepackage{amsfonts}
\usepackage{booktabs}
\usepackage{siunitx}
\usepackage{longtable}
\usepackage{array}
\usepackage{setspace}
\usepackage{float}
\usepackage{nameref} 
\newcommand{\AS}[1]{{\color{black}{#1}}}

\newcommand{\HM}[1]{{\color{black}{#1}}}
\newcolumntype{P}[1]{>{\centering\arraybackslash}p{#1}}
\newcolumntype{M}[1]{>{\centering\arraybackslash}m{#1}}
\usepackage{lineno}
\usepackage[colorlinks,citecolor=red,urlcolor=blue,bookmarks=false,hypertexnames=true]{hyperref}
\begin{document}
\singlespacing

\title{Probing Flow-Induced Chain Scission in Shear-Banding Wormlike Micelles}
\author{Alfredo Scigliani}
\altaffiliation{Present address: Pacific Northwest National Laboratory, Richland, WA 99354, USA.}

\author{Samuel C. Grant}

\author{Hadi Mohammadigoushki}
\email[Corresponding Author:]{hadi.moham@eng.famu.fsu.edu}

\affiliation{Department of Chemical \& Biomedical Engineering, FAMU-FSU College of Engineering, Tallahassee, FL 32310, USA.}
\affiliation{National High Magnetic Field Laboratory, Florida State University, Tallahassee, FL, USA 32310}
\date{\today}

\begin{abstract}
We investigate the microstructural dynamics of wormlike micellar (WLM) solutions based on CPyCl/NaSal/D$_2$O under shear flow using spatially resolved rheo-diffusion NMR, velocity-compensated diffusion measurements, and rheo-optical imaging in a Taylor--Couette geometry. Two viscoelastic systems were examined: a non-shear-banding WLM with a \HM{molar} salt-surfactant concentration ratio of $R=0.4$ and a shear-banding WLM with $R=0.53$. Rheo-optical and velocity measurements confirm homogeneous flow for the non-shear-banding system and the coexistence of low- and high-shear bands within the stress-plateau regime for the shear-banding system. Despite the pronounced heterogeneity in velocity and shear rate associated with shear banding, the local surfactant self-diffusivity remains spatially uniform across the gap and throughout the stress-plateau regime. Furthermore, analysis of the mean-squared-displacement scaling reveals no measurable difference between the high- and low-shear bands. These results demonstrate that the micellar microstructure remains largely unchanged despite the formation of shear bands and provide direct experimental evidence that flow-induced micellar chain scission is not a dominant mechanism governing shear banding in this WLM system.
\end{abstract}
\maketitle

\section{Introduction}
Shear banding phenomena, i.e.,  discontinuities in velocity gradients in flowing materials, have been reported in a host of important systems including granular materials~\cite{kou2017granular, schall2010shear}, emulsions~\cite{paredes2011shear}, suspensions~\cite{ losert2000particle,ovarlez2009phenomenology}, associative polymers~\cite{castillo2017shear, sprakel2008shear}, foam~\cite{kabla2003local} and worm-like micelles (WLMs)~\cite{divoux2016shear}. WLMs or self-assembled surfactant micelles in aqueous solutions are an ideal model for shear banding studies due to their simple rheology, adjustable viscoelastic properties, and shear banding range through parameters like temperature and concentration. A key feature of wormlike micelles is their ability to undergo reversible chain breakage and reformation, which has led to their characterization as living polymers~\cite{cates1987reptation}.\par 

Shear banding has been identified and predicted unambiguously in viscoelastic WLMs~\cite{spenley1996nonmonotonic}, but the microstructural origin of the phenomenon has yet to be fully resolved. Classical Doi-Edwards tube-based models that incorporate the effect of breakage and reformation in chain dynamics suggest that shear banding arises from flow-induced disentanglement of the micellar network~\cite{spenley1993nonlinear,peterson2021predictions}. In these models, the breakage and reformation rates are assumed to be constant and not changed by flows, and therefore, based on these models, shear banding is similar to a first-order phase transition, with aligned micelles in the high shear band and entangled micelles in the low shear band. In this framework, changes in entanglement dynamics under shear are sufficient to produce a nonmonotonic flow curve and trigger band formation in WLMs. Other models attribute shear banding to shear-induced concentration coupling~\cite{schmitt1995shear,fielding2003flow}. \HM{In these models, it is predicted that high shear band to contain smaller  concentration of the micelles than the low shear band.} An alternative theoretical explanation involves flow-induced micellar breakage as a central mechanism. According to this view, increasing shear promotes chain scission, resulting in shorter micelles with faster stress relaxation. This feedback may lead to a decrease in shear stress with increasing shear rate beyond a critical point~\cite{vasquez2007network,zhou2014wormlike,germann2013nonequilibrium,germann2016validation}. The latter models have reproduced some aspects of shear-banding, and other key rheological features observed experimentally~\cite{pipe2010wormlike,mohammadigoushki2019transient,germann2016validation, rassolov2020effects, rassolov2023role}. More recent two-species based models suggest that flow-induced micellar breakage may become important at very high shear rates~\cite{salipante2024two}.\par 

While extensive experimental work has examined shear-induced isotropic–nematic transitions and concentration coupling in concentrated and semi-dilute WLMs~\cite{ cappelaere1997rheology, schmitt1994flow, berret1998structure, helgeson2010direct,gurnon2014spatially}, and flow-induced micellar disentanglement~\cite{Lopez2014} the contribution of flow-induced micellar breakage to shear banding dynamics has, to our knowledge, not yet been directly tested in experiments. Investigating this phenomenon requires a technique capable of spatially resolving micellar length during flow. Typical equilibrium micelle lengths in shear banding WLM systems often exceed 1~$\mu$m~\cite{oelschlaeger2009linear}, posing a limitation for conventional microstructural characterization techniques such as small-angle neutron scattering and small-angle x-ray scattering. These methods are optimized for detecting structures up to a few hundred nanometers~\cite{jensen2014monitoring} and are therefore not ideally suited for resolving the larger-scale micellar structures characteristic of shear banding. \HM{Recently, Cusano et
al.~\cite{cusano2025exploring} investigated a
CPCl/Diclo wormlike micellar fluid and reported that, above a critical Weissenberg number
defined using the equilibrium micellar breakage time, the startup stress response exhibited a
pronounced overshoot followed by a sharp undershoot. The authors attributed this behavior
to flow-induced micellar breakage. Because this behavior occurred at shear rates beyond the shear-banding regime,
they concluded that micellar breakage was independent of shear banding in their system~\cite{cusano2025exploring}.} Nevertheless, direct experimental evaluation of shear-induced micellar breakage in shear banding flows remains a significant challenge in the field.\par 

Recent studies have demonstrated that $^{1}$H diffusion-ordered spectroscopy (DOSY) NMR is sensitive to micellar topology, including linear, spherical, vesicular, and branched structures~\cite{holder2021nuclear, scigliani2023probing, rassolov2022kinetics}. Surfactant self-diffusivity has been shown to scale inversely with micelle contour length, reflecting the slower surfactant self-diffusion in longer micelles~\cite{holder2021nuclear, scigliani2023probing}. Additionally, the surfactant mean-square displacement follows $Z^2(t) \sim T_{\text{diff}}^{\alpha}$, where the exponent $\alpha$ decreases from 1 (for spherical micelles) to 0.5 as micelles elongate into rod-like and wormlike structures, indicating increasingly constrained motion with structural anisotropy~\cite{angelico1998surfactant, ambrosone2001molecular, angelico2012impact,holder2021nuclear, scigliani2023probing}. These observations suggest that surfactant diffusive dynamics may provide a means to probe morphological changes, including potential chain scission, under shear banding conditions.\par

The objective of this study is to investigate the diffusive dynamics of surfactant molecules in shear-banding WLMs and to assess the potential role of flow-induced micellar chain scission in the shear-banding phenomenon. To achieve this goal, we combine spatially and temporally resolved rheo-DOSY NMR measurements with rheo-optical imaging that allows us to simultaneously characterize molecular transport, velocity fields, and flow-induced microstructural changes under shear. The experiments are conducted using CPyCl/NaSal/D$_2$O model WLM systems, which have been extensively studied and are known to exhibit well-defined shear-banding behavior. Measurements are performed in a narrow-gap Taylor--Couette geometry. By comparing the local transport dynamics in homogeneous and shear-banding flows, this study seeks to establish whether the formation of shear bands is accompanied by measurable changes in surfactant diffusion and molecular dynamics.

\section{Materials and Methods}

The WLMs in this study were prepared by dissolving 100 mM Cetylpyridinium Chloride (CPyCl) with varying concentrations of Sodium Salicylate (NaSal) (0-90 mM) in a 0.1 M NaCl brine D$_2$O solution. All reagents were purchased from Sigma-Aldrich, and D$_2$O was obtained from Cambridge Isotopes Laboratories and used as received. Solutions were prepared and stored at $T = 25^\circ\text{C}$ to equilibrate for several days. All experiments were conducted at this temperature.
\HM{D$_2$O was used as the solvent for $^1$H NMR measurements. The proton signal from H$_2$O is very strong and can mask the surfactant peaks. Using a deuterated solvent minimizes solvent interference and improves peak resolution.} \AS{Additionally, the density of D$_2$O (approximately $1.105~\mathrm{g\,cm^{-3}}$ at room temperature), which is higher than that of H$_2$O (approximately $0.997~\mathrm{g\,cm^{-3}}$), was taken into account during sample preparation to ensure accurate concentration calculations. Nevertheless, direct comparisons with concentrations reported for H$_2$O-based systems should be made with caution, as structural and rheological transitions may occur at different nominal concentrations in D$_2$O.}

The rheological properties of the micellar solutions were analyzed using a commercial Anton-Paar MCR 302 rheometer with a standard cone-and-plate geometry. Small amplitude oscillatory shear (SAOS) tests were conducted to measure the elastic and loss moduli across a range of frequencies. A frequency sweep was performed with angular frequencies from 100 rad/s to 0.1 rad/s, collecting 8 data points per decade. From the frequency response, we estimated the longest relaxation time of the micellar solutions by fitting the multi-mode Maxwell model to the SAOS data. Additionally, steady shear experiments were carried out to obtain the quasi-steady flow curves of the wormlike micellar solutions, with shear rates systematically varied from 0.1 $\mathrm{s}^{-1}$ to 50 $\mathrm{s}^{-1}$.

\begin{figure*}[htbp]
\centering
\includegraphics[width=1\linewidth]{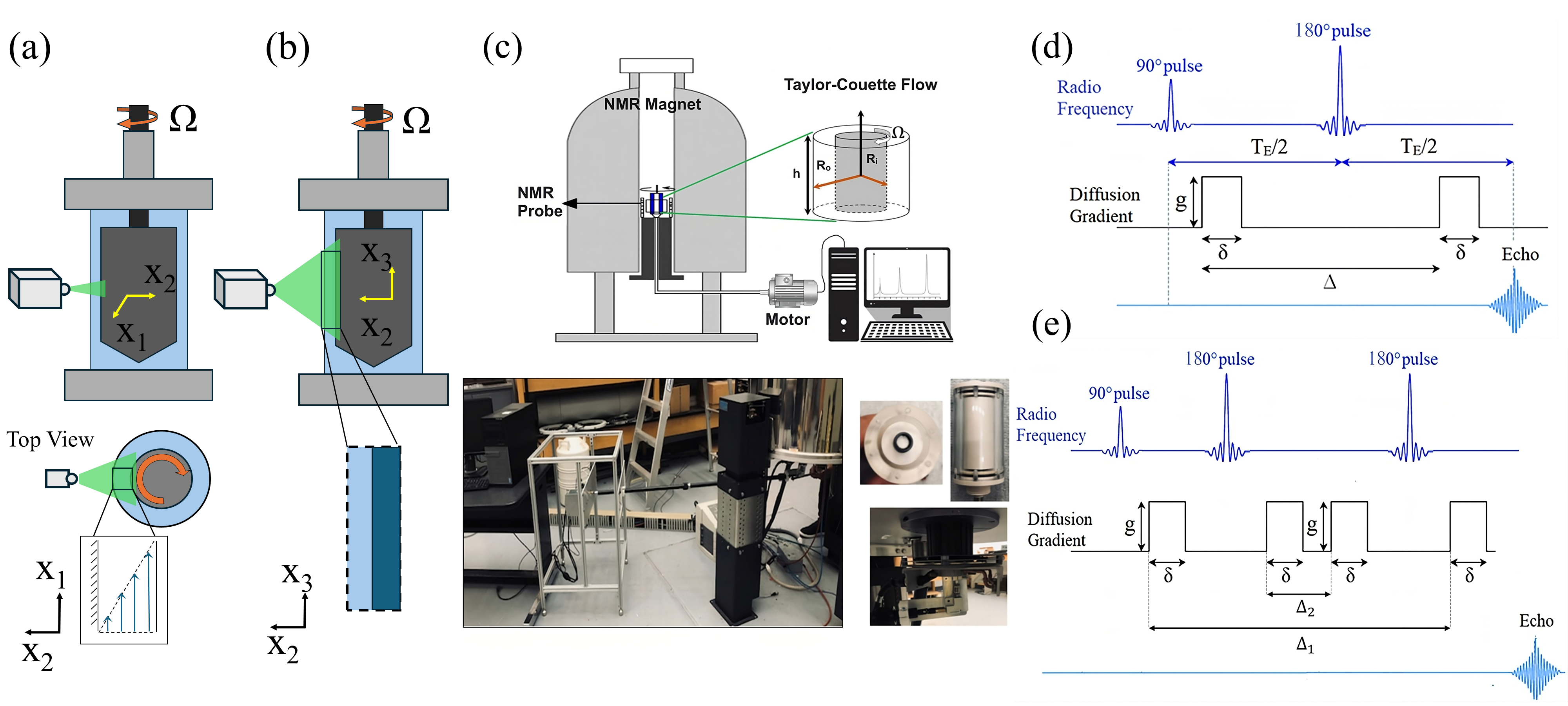}
\caption[Rheo-NMR setup and Sequences]{ A schematic of the rheo-optical cell used to measure the detailed form of flow in the TC cell in the $x_1-x_2$ plane (a), and vorticity gradient $x_2-x_3$ plane (b). (c) A schematic images of the rheo-NMR setup used for experiments under flow. Bottom row: The real images from the rheo-NMR setup with different views of the Taylor-Couette cell. A schematic for the PGSE (d), and the dPGSE (e) sequence. }
\label{fig: RheoNMR Setup}
\end{figure*}

\subsection{Rheo-Optics}
 For the rheo-optical measurements, a small-gap Taylor--Couette (TC) geometry consisting of concentric cylinders with inner and outer radii of $R_i=7.65$ mm and $R_o=8.65$ mm, respectively, was employed. The resulting gap size was $d=1$ mm, and the cylinder height was $H=50$ mm. Two complementary rheo-optical techniques were implemented. First, particle-tracking velocimetry (PTV) was performed in the $x_1$--$x_2$ plane of the TC cell, as illustrated in Fig.~\ref{fig: RheoNMR Setup}(a), to quantify the local velocity field and characterize shear-band formation. The WLM solutions were seeded with glass microspheres (Potters 110P8, diameter $\approx 8,\mu$m) at a concentration of 50 ppm. Particle motion was recorded under laser-sheet illumination and subsequently analyzed to obtain spatially resolved velocity profiles across the gap. Second, rheo-optical turbidity measurements were conducted in the $x_2$--$x_3$ plane of the TC cell, as shown in Fig.~\ref{fig: RheoNMR Setup}(b). In this case, a diverging laser beam was transmitted through the sample while the scattered light intensity was recorded using a digital camera with adjusted gamma settings to enhance contrast.
 
\subsection{Rheo-DOSY-NMR}
$^{1}$H NMR experiments were conducted using an 11.75~T wide-bore magnet at the National High Magnetic Field Laboratory, equipped with a 500-MHz Bruker Avance III spectrometer. The system featured triple-axis Bruker Micro2.5 micro-imaging gradients and Great60 amplifiers, generating a peak 1.5 T/m gradient per axis. Imaging was performed using ParaVision 6.0.1 with customized pulse sequences. A 25-mm linear birdcage probe was used inside a custom-built (TC) cell with a similar geometry to that used in the rheo-optical measurements. A schematic and the real images of the TC cell placed in the NMR magnet are shown in Fig.~\ref{fig: RheoNMR Setup}(c).  The flow was imposed by an STM23Q-3RN NEMA-23 stepper motor (Applied Motion Products). A typical NMR sequence that has been used to assess the velocity profile of fluids is a Pulsed Gradient Spin Echo (PGSE) sequence (see Fig.~\ref{fig: RheoNMR Setup}(d) for sequence details), and for measuring diffusion under flow a double-pulsed gradient spin echo (dPGSE) is used (see Fig.~\ref{fig: RheoNMR Setup}(e) for sequence details). The diffusion encoding parameters were $\delta = 6$ ms, $\Delta = 22$ ms, $T_E = 30$ ms for PGSE and  $T_E = 42$ ms for dPGSE,  with gradient amplitude $g$ varying from 0 to 150 G/cm based on experimental conditions.  

\subsubsection{One-Dimensional Velocity Measurements Using Standard PGSE NMR}

To measure the temporally averaged 1D velocity profiles within the gap of the TC cell, we used a method known as dynamic q-space imaging via the PGSE sequence shown in Fig.~\ref{fig: RheoNMR Setup}(c)~\cite{callaghan1991velocity}. Fig.~S1 shows a schematic step-by-step process. In this process, initially, a time-domain Free Induction Decay (FID) signal is acquired in a spatially-resolved manner within the gap of the TC cell. Next, Fourier transforms are applied to this FID signal, and generate real and imaginary filtered data sets. These data sets are then used for image reconstruction, representing the real and imaginary components of the signal. Each image corresponds to a slice in the q-space and is used to build real and imaginary q-space FIDs. Further analysis involves applying a Fourier transform to the q-space FIDs, resulting in a dynamic displacement profile denoted in Fig.~S1(c,f), with the horizontal axis indicating the number of points in Z space. Ultimately, the final step includes the calculation of velocity images based on the acquired data. The derivation of the equations for $v$ begins with the analysis of the dynamic displacement profile (DDP), which is a Gaussian function centered at the displacement $v$ as shown in Fig.~S1(c,f). Under discrete sampling, the digital position of the DDP peak ($k_v$) is directly proportional to the physical velocity. The relation is given by:
\begin{equation}
k_v = \frac{v N \gamma \delta \Delta g_{\text{max}}}{2\pi M},
\end{equation}

Where N represents the total number of points in the acquired FID (N=2000 was used in our experiments), $\gamma=2.675\times10^8$ rad/(sT), the gyromagnetic ratio of the proton, $g_{max}$ represents the maximum applied diffusion gradient pulse strength (for experiments with WLMs $g_{max}=62.5$ G/cm was used, whereas $g_{max}=20.0$ G/cm for calibration experiments with water samples), and $M$ represents the magnitude of NMR signal intensity. Note that the time required to obtain a good signal-to-noise ratio is 30-60 minutes for each imposed diffusion weighting value, depending on the concentration and the imposed shear rate. 

\subsubsection{Velocity-Compensated NMR Measurements of Diffusion Under Flow}

To measure the molecular diffusivity under strong flows, the velocity-compensated NMR was carried out using a double-pulsed gradient spin echo (dPGSE) shown in Fig.~\ref{fig: RheoNMR Setup}(e). This method was originally developed for flow in a capillary tube~\cite{callaghan1991velocity} but modified by us in this study for a TC cell geometry. The dPGSE sequence shown in Fig.~\ref{fig: RheoNMR Setup}(e) includes an additional pair of gradient pulses applied in the middle of the standard PGSE sequence. These extra gradient pulses have the same magnitude as the initial ones. The advantage of using dPGSE lies in its ability to nullify the effects of coherent motion while preserving the information related to incoherence (or molecular diffusion). Note that for dPGSE, $\Delta = \Delta_1 - \Delta_2$ with $\Delta_1 = {30}$ ms, and $\Delta_2 = {8} $ ms.  \\

\begin{figure*}[ht]
\centering
\includegraphics[width=1\linewidth]{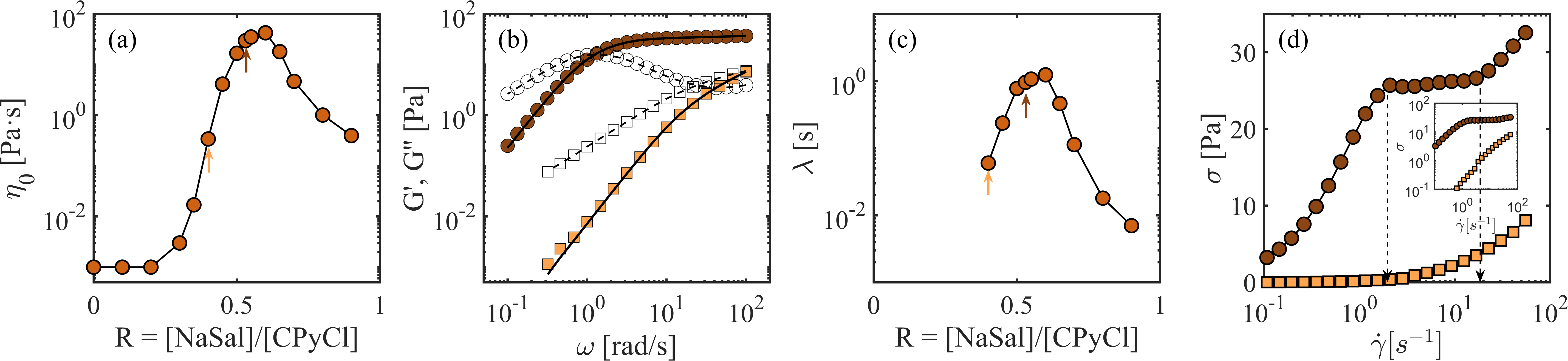}
\caption{(a) Zero shear viscosity ($\eta_0$) as a function of salt to surfactant concentration for the micellar solution based on CPyCl/NaSal in D$_2$O. (b) Longest relaxation time as a function of salt to surfactant concentration ratio $R$ for various micellar solutions. The arrows in (a,b) denote the weakly viscoelastic WLMs with $R = 0.4$ and shear banding WLMs with $R = 0.53$. (c) Steady-state shear stress as a function of imposed shear rate for $R=0.4$ (squares), and for $R = 0.53$ (circles). The vertical arrows denote the onset $\dot{\gamma}_l\approx1.7~\mathrm{s}^{-1}$ and the end $\dot{\gamma}_h\approx18~\mathrm{s}^{-1}$ of the stress plateau. (d) Small amplitude oscillatory shear experiments as a function of angular frequency for the WLMs with $R = 0.4$ (squares), and shear banding WLMs with $R = 0.53$ (circles). The longest relaxation time was obtained by fitting data to a multi-mode Maxwell model (shown as dashed and continuous curves), and numerical values are $\lambda = 0.07$s for $R=0.4$ and $\lambda = 1$s for $R=0.53$.  }
\label{fig: bulk_rheology}
\end{figure*}
\section{Fluid Selection and Characterization}
The primary objective of this study is to quantify surfactant self-diffusive dynamics in shear-banding flows using Rheo-NMR. To this end, we will select three model fluids spanning increasing levels of microstructural complexity. First, a Newtonian fluid (de-ionized water) will be used to calibrate and validate the Rheo-NMR methodology under well-characterized flow conditions. Second, a wormlike micellar solution composed of relatively short micelles will be chosen as a non-shear-banding reference system that exhibits a monotonic flow curve without a stress plateau. Finally, a wormlike micellar solution known to undergo shear banding will be selected. The latter system should exhibit a characteristic stress plateau. \par 

To accomplish this objective, the surfactant concentration was held constant at 100 mM while the salt-to-surfactant molar ratio was varied over a broad range ($R = 0$-$0.9$). Figures~\ref{fig: bulk_rheology}(a,b) present the measured zero-shear viscosity and longest relaxation time of the resulting wormlike micellar solutions. As $R$ increases, the zero-shear viscosity initially rises and reaches a local maximum, which has been attributed to the transition from predominantly linear micelles to branched micellar structures in this system~\cite{gaudino2015adding,omidvar2019detecting,holder2021nuclear}. The longest relaxation time exhibits a similar dependence on salt concentration, which reflects a similar underlying evolution of the micellar microstructure. To establish a non-shear-banding reference fluid, we selected the solution with $R=0.4$. Although this fluid is viscoelastic, its flow curve remains monotonic and exhibits no evidence of a stress plateau or shear banding, as shown in Fig.~\ref{fig: bulk_rheology}(c,d). At this salt composition, the micelles are relatively short, do not exhibit shear banding, and are unlikely to undergo significant shear-induced breakage in the imposed shear-rate range, thus serving as a benchmark to isolate shear-banding effects. Increasing the salt concentration beyond $R=0.4$ promotes micellar growth and ultimately leads to the onset of shear banding. Furthermore, we selected the solution with $R=0.53$ as the shear-banding system. This sample exhibits a pronounced stress plateau in its flow curve as shown in Fig.~\ref{fig: bulk_rheology}(c). In addition, this shear banding wormlike micelle is within the fast-breaking regime, and therefore, it is likely to undergo shear-induced micellar breakage. The ratio of breakage time to reptation time for the shear banding wormlike micelle sample is estimated using correlations proposed by Larson and co-workers~\cite{tan2022quantitative}. The reptation time ($\tau_{rep}$) of the micelle chains could be estimated as:
\begin{equation}
    \tau_{rep} = \frac{2 {\langle L\rangle}^3}{\pi^2 \chi D_o},
\end{equation}

where $\langle L\rangle = l_e Z$ is the average micelle length, $\chi=l_e/l_p$ is the ratio of the entanglement length to the persistence length, and $D_0=\frac{\ln(\xi/d)k_BT}{2\pi\eta_s}$ is the micellar translational diffusivity within the tube~\cite{tan2022quantitative}. Here, $\xi$ is the mesh size, $d$ is the micelle diameter, $k_B$ is the Boltzmann constant, $T$ is the temperature, and $\eta_s \approx 1.1\times10^{-3}$~Pa$\cdot$s is the solvent viscosity. The mesh size is estimated from the plateau modulus using $G_0=k_BT/\xi^3$. The micelles entanglement density $Z^{0.82} = {3.15 G_o/G''_{min}}$. Therefore, the micelles entanglement length is evaluated to be $l_e=\xi^{5/3}/l_p^{2/3}$. We estimate the persistence length as $l_p\approx28$~nm based on experimental data reported for CPyCl/NaSal solutions in~\cite{oelschlaeger2009linear}. Once $\tau_{\mathrm{rep}}$ is determined, the micellar breakage time $\tau_{\mathrm{br}}$ is obtained from $\lambda=\sqrt{\tau_{\mathrm{br}}\tau_{\mathrm{rep}}}$. Here, $\lambda$ is the longest relaxation time, obtained by fitting the Maxwell model to the SAOS data shown in Fig.~\ref{fig: bulk_rheology}(d). The estimated parameters are summarized in the table~(\ref{Table:Rheology}) below. The resulting $\zeta\ll1$ indicates that the micellar system is in the fast-breaking regime. Note that the persistence length and micelle diameter used to estimate $\zeta$ were taken from measurements in H$_2$O-based micellar solutions. Therefore, the estimated $\zeta$ should be considered approximate, as replacing H$_2$O with D$_2$O may alter the micellar structure and dimensions.\\

\begin{table}[H]
    \centering
    \begin{tabular}{c@{\hspace{0.1cm}}c@{\hspace{0.1cm}}c@{\hspace{0.1cm}}c@{\hspace{0.1cm}}c@{\hspace{0.1cm}}c@{\hspace{0.1cm}}c}
        \hline
         $G_o$ [Pa]& $\chi$& $d$ [nm] & $D_o$ [m$^2$/s] &  $\tau_{rep}$ [s] & $\tau_{br}$  [s]& $\zeta ( =\tau_{br}/\tau_{rep})$ \\ \hline
   
         35&3.8 & $\approx$ 5~\cite{raghavan2017wormlike}& 1.5$\times 10^{-18}$ & 6.7 & 0.14 & 2.2$\times10^{-2}$ \\ \hline
    \end{tabular}
    \caption{A summary of the microstructural and rheological properties of the shear banding wormlike micellar system (with $R=0.53$) selected for this study.}
    \label{Table:Rheology}
\end{table}

\section{Results and Discussion}
\begin{figure*}[htbp]
\centering
\includegraphics[width=1\linewidth]{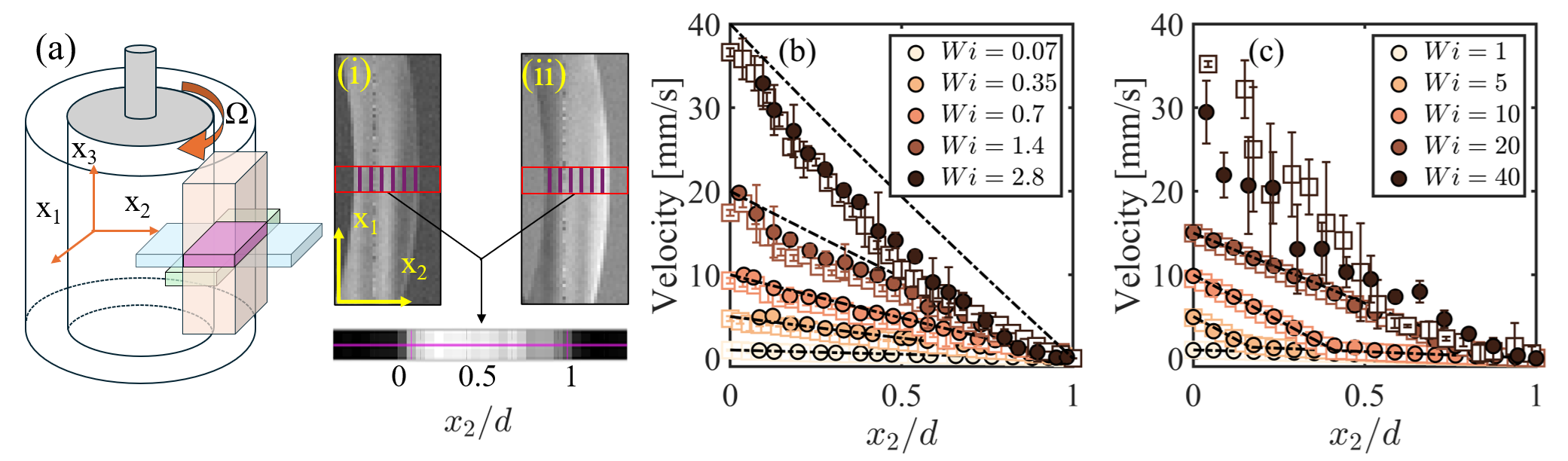}
\caption{(a) Schematic of the Taylor–Couette cell showing three overlapping slices forming the region of interest across the gap. Reconstructed images of the real (i) and imaginary (ii) NMR signal components are used to obtain the local velocity field. 1D time-averaged velocity profiles for weakly viscoelastic WLMs (b) and shear-banding WLMs (c) measured by rheo-optics (squares) and PGSE-NMR (circles). Dashed and Black lines in (c,d) show predictions from the Newtonian behavior and lever rule, respectively.}
\label{fig: PTV_NMR_velocity}
\end{figure*}
Fig.~\ref{fig: PTV_NMR_velocity}(a) presents a schematic of the TC geometry together with the three-directional frequency encoding scheme used in this study to acquire time-resolved MR images within the flow plane ($x_1$-$x_2$) of the TC cell. Under flow, the NMR signal attenuation comprises two distinct contributions: an incoherent term arising from random molecular diffusion and a coherent term originating from bulk advection with velocity 
$v$ and is given as~\cite{carr1954effects,callaghan1991velocity}:

\begin{equation}
\frac{S(q)}{S_0} = \exp\left[-q^2 T_{\mathrm{diff}} D + i q T_{\mathrm{diff}} v \right].
\label{attenuationflow}
\end{equation}Here $D$ is the self-diffusion coefficient of the molecule, $T_{\mathrm{diff}}$ is the diffusion delay time with  $T_{\mathrm{diff}} = \Delta-\delta/3$. Additionally, $S(q)$ and $S_0$ denote the NMR signal intensities acquired with and without diffusion weighting, respectively. In addition, the diffusion weighting parameter $q = \gamma^2 \delta^2 g^2$. In the following, we first determine the local velocity field to confirm the moderate shear-thinning and shear-banding nature of the wormlike micellar systems, as well as the profile for Newtonian de-ionized water. We then extract the local surfactant diffusivity under flow.\par

The signal attenuation of Eq.~(\ref{attenuationflow}) is produced through both molecular diffusion and advective motion. For a meaningful determination of either velocity or diffusivity, one should make sure to decouple the signal attenuation associated with each of these contributions. It is therefore essential to quantify the relative contribution of each factor in overall signal attenuation. To accomplish this, we compare the characteristic spin displacement due to flow over the diffusion encoding time $T_{\mathrm{diff}}$, namely $v T_{\mathrm{diff}}$, with the characteristic diffusive displacement $\sqrt{D T_{\mathrm{diff}}}$. Their ratio defines a dimensionless Peclet number as:
\begin{equation}
   \mathrm{Pe} =  \frac{v T_{\mathrm{diff}}}{\sqrt{D T_{\mathrm{diff}}}}
    = \frac{v}{\sqrt{D/T_{\mathrm{diff}}}}.
\end{equation}

Because the velocity field in the Taylor--Couette gap is spatially non-uniform, we adopt a conservative estimate of the above dimensionless number by using the maximum velocity within the TC cell. The maximum velocity occurs at the inner cylinder and scales as $ v_{\max} = \dot{\gamma} d$, where $\dot{\gamma}$ is the imposed shear rate. Substituting this into the above expression yields:
\begin{equation}
    \mathrm{Pe} = \frac{\dot{\gamma} d}{\sqrt{D/T_{\mathrm{diff}}}}.
\end{equation}

This dimensionless group provides an estimate of the relative importance of advective and diffusive displacements during the encoding time. When $\mathrm{Pe} \gg 1$, signal attenuation is dominated by flow; when $\mathrm{Pe} \ll 1$, diffusion dominates. For the WLM systems studied here, $\mathrm{Pe}> 1$ for $\dot{\gamma}> 0.1~\mathrm{s}^{-1}$, indicating that signal attenuation is dominated by the advective term over the entire shear-rate range considered in this study (1–40 $\mathrm{s}^{-1}$). Consequently, the 1D velocity field can be measured using a standard pulse-gradient spin-echo (PGSE) sequence.\par

\subsection{Quantifying the Details of Flow Field}

Representative quasi-steady one-dimensional velocity profiles obtained by PGSE and an independent rheo-optics method are shown in Fig.~\ref{fig: PTV_NMR_velocity}(b,c). Corresponding measurements for deionized water are also provided in the Fig.~S2 of the supplementary materials. For $R=0.4$, the profiles are nearly Newtonian (which is shown as a dashed line in Fig.~\ref{fig: PTV_NMR_velocity}(b)) at low shear rates and progressively deviate at higher shear rates from Newtonian behavior, consistent with their bulk viscosity, which shows a minimal shear thinning at low shear rates and moderate shear thinning at higher shear rates. In contrast, the $R=0.53$ system exhibits a clear shear-banded velocity profile for shear rates within the stress plateau, consistent with the lever-rule predictions (illustrated by dashed lines in Fig.~\ref{fig: PTV_NMR_velocity}(c)). At shear rates above the stress plateau (e.g., $\dot{\gamma} = $ 40 $\mathrm{s}^{-1}$), the 1D time-averaged velocity field strongly fluctuates and exhibits spatial heterogeneity, which is reminiscent of viscoelastic instabilities and elastic turbulence (we will return to this feature below). Overall, the 1D averaged velocities obtained via PGSE are consistent with those obtained by rheo-optics for both wormlike micellar systems. \HM{The individual velocity profiles obtained by either method are plotted separately as Fig.~S3 in the supplementary materials. } \par 

\begin{figure*}[ht]
\centering
\includegraphics[width=1\linewidth]{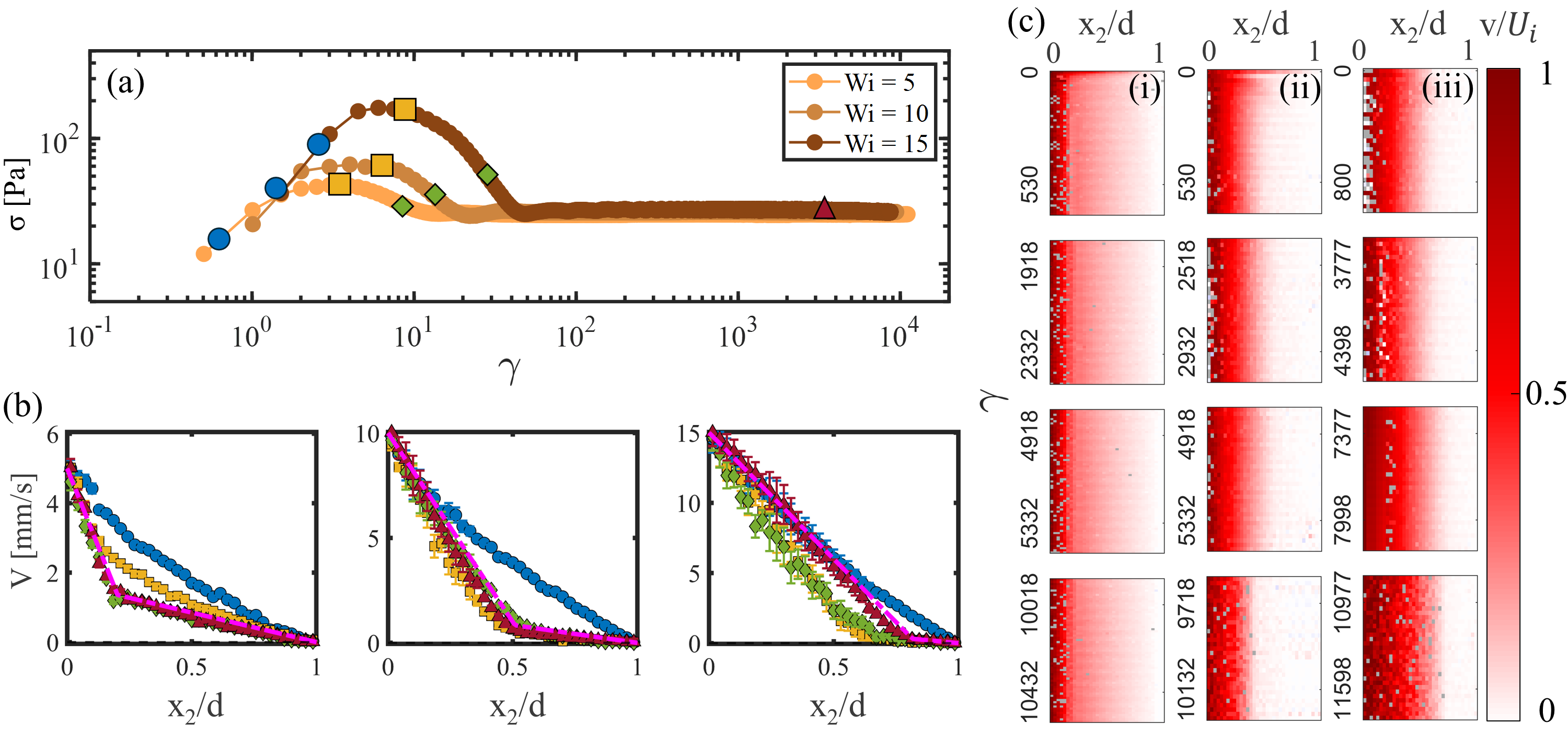}
\caption{(a) Temporal evolution of shear stress for multiple imposed shear rates. (b) Temporal evolution of 1D velocity profiles measured for the shear banding WLMs in the $x_1-x_2$ plane at (i) $Wi = 5$, (ii) $Wi = 10~ \mathrm{s^{-1}}$, and (iii) $Wi = 15 $ compared with predictions of the lever-rule for quasi-steady velocity profiles (magenta dotted lines). (c) spatio-temporal evolution of velocity profiles for various imposed shear rates from left to right for (i) $Wi = 5 $, (ii) $Wi = 10 $, and (iii) $Wi = 15 $ in the $x_1-x_2$ plane. Here U$_i$ and $\gamma$ denote the imposed velocity of the inner cylinder and the strain defined as $\gamma = Wit$.}
\label{fig: 2D_PTV}
\end{figure*}

{On the other hand, extracting molecular diffusivity under flow is challenging because, as shown before, NMR signal attenuation is dominated by coherent phase dispersion arising from bulk flow for $\dot{\gamma}\gtrsim 0.1~\mathrm{s}^{-1}$. To accurately quantify diffusion under flow, the advective contribution to NMR signal attenuation must be separated from the diffusive contribution. This separation is possible because advection and diffusion affect the NMR phase attenuation differently: steady flow produces coherent phase shifts, whereas diffusion generates random phase dispersion~\cite{callaghan1991velocity,carr1954effects}. Under steady flows, the phase dispersion due to bulk motion can therefore be refocused using appropriately designed radio-frequency pulse sequences (see Fig.~\ref{fig: RheoNMR Setup}(e)). In essence, the dPGSE sequence shown in Fig.~\ref{fig: RheoNMR Setup}(e) achieves this by employing two symmetric pairs of diffusion gradient pulses separated by the same diffusion time $\Delta$, with a 180$^{\circ}$ refocusing pulse between them. The third 180$^{\circ}$ refocusing pulse inverses the direction of spins, and the similarity between the second pair of diffusion gradients and those of the first one ensures that phase shifts acquired by spins moving coherently with the flow during the first gradient pair are precisely reversed by the second pair. This enables an effective cancellation of the bulk flow’s contribution to the net spin displacement and consequently the net phase of the NMR signal. As a result, although the sample continues to flow with its natural velocity profile, the coherent motion’s effect on the measured NMR signal attenuation can be eliminated in theory in this way. This approach, known as velocity-compensated NMR diffusometry using a double pulsed-gradient spin-echo (dPGSE) sequence, was first noted by Carr and Purcell~\cite{carr1954effects} and later developed and validated for Newtonian capillary flows by Callaghan and co-workers~\cite{callaghan1991velocity}.}

{The key requirement for this approach is that the bulk flow should be steady. Shear-banding flows are susceptible to viscoelastic instabilities that can drive spatio-temporal fluctuations of the high-shear band~\cite{lerouge2008interface,lerouge2006interface,mohammadigoushki2017inertio,scigliani2025study,mohammadigoushki2016flow}. It is therefore essential to assess the presence of such heterogeneities in the present WLM systems before we attempt the velocity-compensated NMR measurements. For the non-shear banding wormlike micelle with $R=0.4$, the flow remains spatially uniform up to the highest imposed shear rate, $\dot{\gamma}=40 \mathrm{s}^{-1}$ (see Movie S1 for the flow in the vorticity plane of the TC cell), and no flow instabilities are observed. For the shear-banding system, we measured the temporal evolution of 1D velocity profiles across the TC gap, as shown in Fig.~(\ref{fig: 2D_PTV}). For imposed shear rates within the stress-plateau regime, the velocity profiles remained invariant over time, and the interface separating the high and low-shear bands remained stationary. No measurable interface migration or significant temporal fluctuations were observed during the experiments, indicating that the shear-banded state is temporally stable on the timescale of the measurements \AS{(See Fig.~S4 of the supplementary materials)}. In contrast, for imposed shear rates beyond the stress-plateau regime, the velocity field exhibited pronounced temporal fluctuations and irregular dynamics. Characterizing the temporal evolution of these velocity profiles became increasingly challenging due to the emergence of highly three-dimensional flow structures associated with elastic turbulence. These observations demonstrate a clear transition from a temporally stable shear-banded state within the stress plateau to a strongly unsteady flow state at higher shear rates.\par 

To further verify the stability of the high-shear band throughout the stress-plateau regime, we visualized the vorticity plane of the TC cell. These measurements were performed to assess the presence of any secondary flow structures or spatiotemporal instabilities that may not be evident from 1D velocity profiles alone and to confirm that the interface between high and low-shear bands remains stationary across the stress-plateau range. Some WLM solutions exhibit distinct turbidity contrasts between the high- and low-shear bands in the TC cell's $x_2$--$x_3$ plane~\cite{fardin2012shear,mohammadigoushki2016flow}. As a result, measurements of the turbidity field and its spatio-temporal evolution have become a powerful diagnostic tool for characterizing shear-banding dynamics and identifying flow instabilities in WLM systems. Fig.~\ref{fig: turbidity}(a) shows the spatio-temporal evolution of turbidity in the vorticity ($x_2-x_3$) plane of the TC cell for the shear-banding system ($R=0.53$) following shear startup at $\dot{\gamma} = 15~\mathrm{s}^{-1}$, which is the highest tested shear rate within the stress plateau. 

\begin{figure*}[htbp]
\centering
\includegraphics[width=1\linewidth]{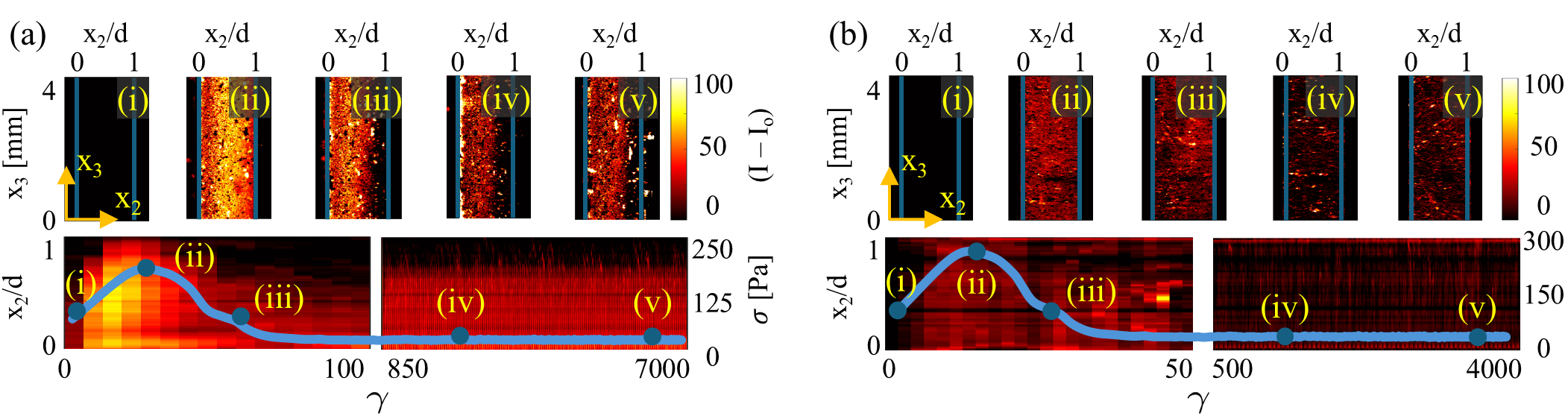}
\caption{Turbidity visualized by illuminating the $x_2-x_3$ plane for the shear banding WLMs at Wi = 15 (a), and Wi = 40 (b) at various strain values ($\gamma$) following the inception of step shear. Strain is defined as $\gamma = \dot{\gamma}t$. Blue curves show the temporal evolution of the shear stress.}
\label{fig: turbidity}
\end{figure*}

Shortly after imposition of flow, a turbid band nucleates near the inner cylinder, spans the gap at the stress overshoot, and subsequently retracts to form a sharp interface between high (bright) and low-shear (dark) bands. At long times, as the stress approaches steady state, two distinct bands form, with the high-shear band occupying $\approx 75\%$ of the gap (consistent with Fig.~\ref{fig: PTV_NMR_velocity}(c)). Similar results for lower shear rates are shown in Fig.~S5 of the supplementary materials. \HM{We verified the long-term stability of the shear-banded interface by tracking its position across a wide range of accumulated strains. For applied shear rates within the stress plateau region, the boundary separating the high and low shear bands rapidly stabilizes. \AS{As shown in Fig.~S6 of the supporting information, the interface location obtained from turbidity measurements remains virtually stationary throughout the shear duration consistent with interface locations obtained reported in Fig.~S4.} It exhibits negligible spatial drift even at high strain values. This result confirms that the shear-banded profile achieves a time-independent steady state, ruling out transient interfacial migration  under sustained deformation.} At shear rates above the stress plateau (e.g., $Wi=20$ and $40$), a chaotic behavior is observed in turbidity even after long times, which is indicative of elastic instabilities (as shown in Fig.~\ref{fig: turbidity}(b) and Movie S1). Based on the above analysis, velocity-compensated NMR diffusometry can reliably measure self-diffusivity within the stress plateau, where the flow remains quasi-steady. Those results will be reported below. \par}

\subsection{Quantifying Molecular Diffusion Under Flow}
\begin{figure*}[htbp]
\centering
\includegraphics[width=1\linewidth]{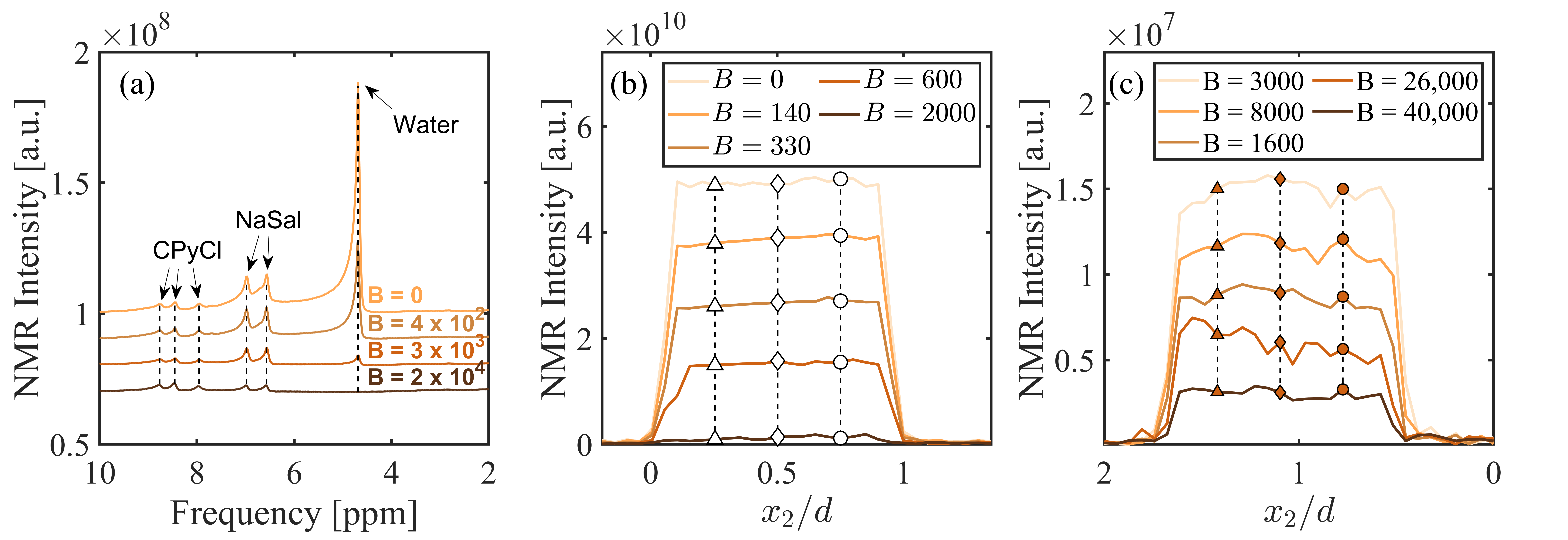}
\caption{(a) Bulk $^1$H NMR spectra of shear-banding WLMs in the TC cell acquired at increasing diffusion weighting, $B$. (b) Spatially resolved velocity-compensated NMR signal intensity for water (b), and surfactants (c) in shear-banding WLMs as a function of $B$ at $Wi=15$. }
\label{fig: spectra_magnitude}
\end{figure*}

Subsequently, we performed velocity-compensated rheo-DOSY NMR experiments to quantify the spatial evolution of surfactant self-diffusion under shear flow for imposed shear rates within the stress plateau. A critical prerequisite for these measurements is the unambiguous identification of the surfactant resonances and their separation from signals that arise from other molecules that are present in the wormlike micellar solution. Figure~\ref{fig: spectra_magnitude}(a) presents the bulk $^1\text{H}$ NMR spectra of the shear-banding WLM measured across the entire TC cell, which reveals resonances associated with the different species that exist in the wormlike micellar solution. The spectrum obtained for the non-shear-banding system was qualitatively similar and is therefore not shown. The dominant resonance at 4.7 ppm originates from residual water protons in $\mathrm{D_2O}$, whereas the remaining well-defined peaks are assigned to CPyCl and the associated counter-ion NaSal, which is bound to CPyCl. Importantly, the surfactant resonances are clearly separated from the water peak, which enables selective and reliable characterization of surfactant self-diffusion. Another difference between the water peak and those of surfactants is the rate by which the signals attenuate, as shown in Fig.~\ref{fig: spectra_magnitude}(a). As the diffusion weighting ($B$) increases, the water resonance rapidly attenuates. Here $B$ is defined as $B = q^2 T_{\mathrm{diff}}$. In contrast, the signal decay for the larger surfactant molecules and their bound counter-ions is significantly slower, consistent with previous measurements on similar systems~\cite{holder2021nuclear, scigliani2023probing}. 

Therefore, a combination of frequency-selective excitation and strong diffusion weighting enables robust quantification of surfactant self-diffusion with negligible interference from the solvent contribution. Figure~\ref{fig: spectra_magnitude}(b,c) present the spatially resolved NMR signal intensity of water and surfactant molecules across the TC gap over a broad range of diffusion weightings, $B$ and at an imposed shear rate of $\dot{\gamma} = 15~ s^{-1}$. As the diffusion weighting $B$ increases, the signal intensity of both species across the gap of the TC cell remains uniform but decreases due to diffusion-induced attenuation. However, the attenuation rates differ significantly. As shown in Fig.~\ref{fig: spectra_magnitude}(b), the water signal decays much more rapidly than the surfactant signal and becomes comparable to the noise level at $B\approx 2\times 10^3~ \text{s/mm}^2$. In contrast, the surfactant signal attenuates considerably more slowly and remains measurable even at $B\approx 4\times 10^4 ~\text{s/mm}^2$. \HM{This result is expected because surfactant molecules are bound to the micelles, and therefore, experience a much slower diffusion process than those experienced by the water molecule. }\par 

\begin{figure*}[t!]
\centering
\includegraphics[width=1\linewidth]{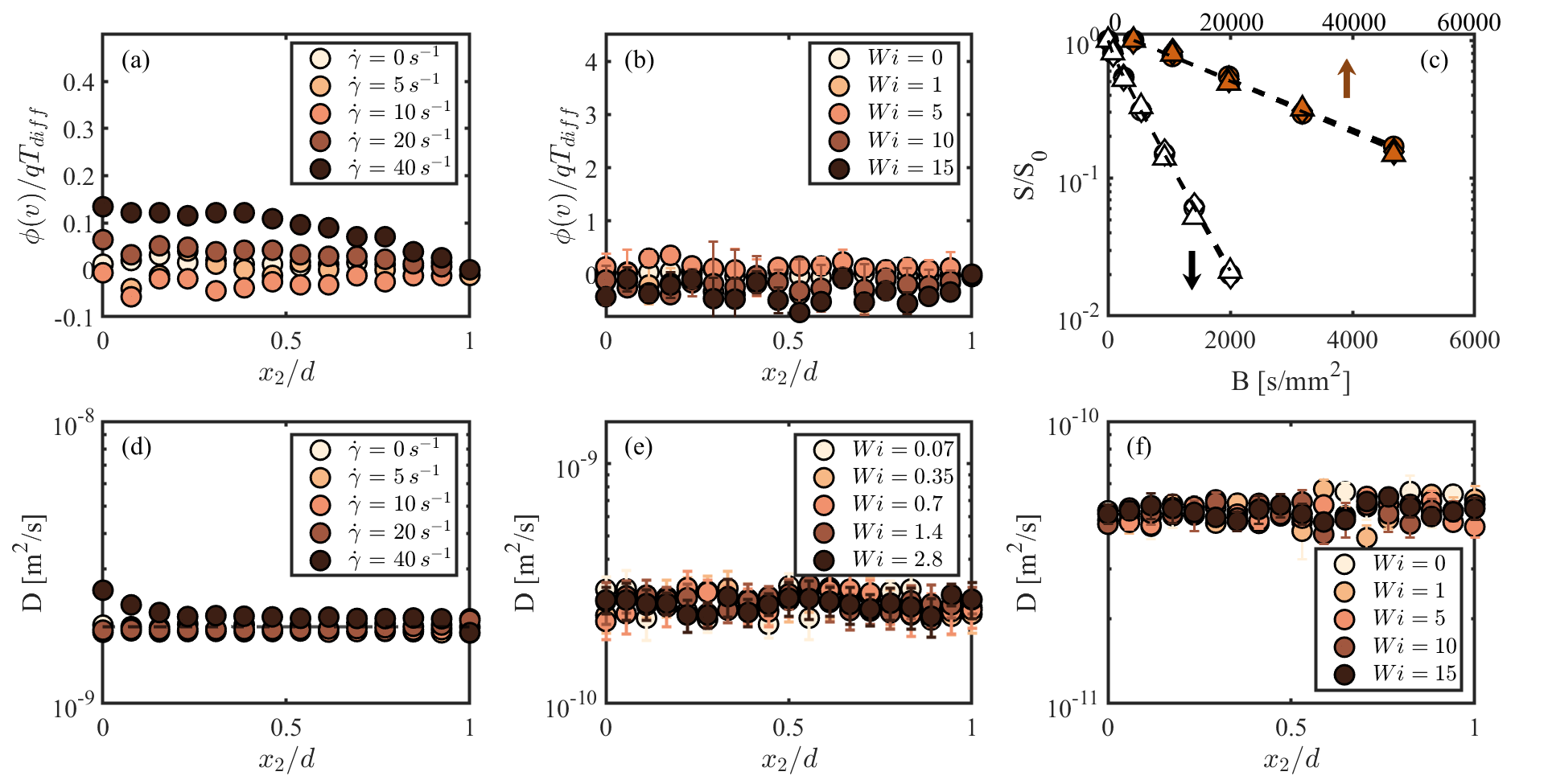}
\caption{Velocity-compensated signal attenuation arising from advective flow ($\phi (v)= qT_\mathrm{diff}v$) across the TC gap at various imposed shear rates for water (a), and surfactant molecules in WLMs with $R=0.53$ (b). (c) Signal attenuation as a function of diffusion weighting $B$ for water (empty symbols) and surfactants (filled symbols). The dashed lines are the best fit of Eq.~\ref{attenuationflow} to data. Diffusion coefficient measured over the gap of the TC cell for residual protons in solvent (d), and surfactant in the wormlike micelles with $R=0.4$ (e), and $R=0.53$ (f) over a broad range of imposed shear rates.}
\label{fig: dPGSE_velocity}
\end{figure*}

To quantify the contribution of advection to the measured NMR signal, Fig.~(\ref{fig: dPGSE_velocity}) presents the residual phase shift arising from bulk flow across the TC gap, expressed as $\phi(v)/qT_{\mathrm{diff}}$ for residual water in the D$_2$O solvent in Fig.~\ref{fig: dPGSE_velocity}(a) and for the surfactant molecules in Fig.~\ref{fig: dPGSE_velocity}(b). \HM{Note that the residual phase shift is effectively a measure of the local fluid velocity under flow. For water, the compensated velocity remains close to zero at all shear rates, indicating that the velocity contribution is largely removed. Small deviations from zero appear at the highest shear rates, which will be discussed later. In contrast, the shear-banding WLM solutions exhibit fluctuations around zero, with $\phi(v)/qT_{\rm diff}=0\pm0.5$. However, these fluctuations remain small compared to the imposed flow velocity. For example, at $Wi=5$, the inner-cylinder velocity is approximately 5 mm/s, whereas the residual variations are less than 10\% of this value. As $Wi$ increases, the imposed velocity increases while the magnitude of the residual variations remains nearly unchanged, causing the relative deviation of velocity to decrease to well below 10\%. } 

Therefore, for all imposed $Wi$ values, this residual attenuation remains within experimental noise, which demonstrate successful suppression of flow-induced phase dispersion by the velocity-compensated method. Similar results for surfactants in the WLMs with $R = 0.4$ are reported in Fig.~S7 of the supplementary materials. Thus, the attenuation in Fig.~\ref{fig: dPGSE_velocity}(b) reflects only incoherent surfactant diffusivity. Fig.~\ref{fig: dPGSE_velocity}(c) shows the averaged signal attenuation at selected positions denoted in Fig.~\ref{fig: spectra_magnitude}(b,c) within the TC gap and the corresponding fits to Eq.~(\ref{attenuationflow}). \HM{Notably, the signal decay for the residual water peak is significantly faster than that of the surfactant species. This is expected, as surfactant molecules diffuse much more slowly in the micelles than free water in the solvent. Furthermore, the surfactant signal attenuation follows a mono-exponential decay at high $B$-values. } Fitting the signal decay to this equation allows us to evaluate the diffusion coefficient ($D$) of species across the TC gap under different flow conditions. \par
Figure~(\ref{fig: dPGSE_velocity}) presents the spatially resolved self-diffusion coefficients of residual water in D$_2$O (d) and surfactant molecules in wormlike micellar solutions with $R=0.4$ (e) and $R=0.53$ (f). As expected, the self-diffusion coefficient of residual water in D$_2$O solvent remains essentially constant across the TC gap and is independent of the imposed shear rate. In addition, these data are similar to the diffusion coefficient of water that is obtained through an independent bulk DOSY experiments with $D = 2.2\times 10^{-9}$ m$^2$/s. This observation is consistent with the fact that the solvent is not expected to undergo significant structural changes under shear and provides further validation that the velocity-compensated diffusion measurements accurately isolate molecular diffusion from convective motion.

 \HM{However, at the highest imposed shear rate, $\dot{\gamma}=40~\mathrm{s^{-1}}$, the diffusion coefficient of water increases by approximately 20\%. In particular, at the inner cylinder, the residual water diffusivity increases beyond the expected diffusion coefficient. To assess whether this increase is associated with inertial flow instabilities, we estimated the Taylor number,
\begin{equation}Ta =  Re\sqrt{\frac{d}{R_i}},
\end{equation}
where $Re$ is the Reynolds number, $d$ is the gap width, and $R_i$ is the inner-cylinder radius. Based on the Taylor–Couette geometry and $\text{D}_2\text{O}$ solvent properties, the Taylor number at the maximum shear rate ($\dot{\gamma} = 40~ \mathrm{s^{-1}}$) reaches $Ta \approx 40$, approaching the critical threshold for Taylor vortex instabilities ($Ta_c \approx 41.2$). The observed increase in apparent diffusivity is therefore likely driven by bulk inertial instabilities. Close to this transition, secondary flows induced by end-wall effects and finite cell height may contribute to this enhanced mixing and unsteady flow. This behavior highlights the theoretical limits of velocity-compensated NMR diffusometry noted earlier in the paper. While velocity compensation successfully removes phase shifts under steady, unidirectional laminar flow, it does not account for velocity fluctuations or non-laminar streamlines near an unsteady transition. Consequently, uncompensated phase dispersion caused by these secondary and inertial flows manifests as an apparent enhancement in the diffusion coefficient. }\par
\begin{figure}[ht]
\centering
\includegraphics[width=0.50\linewidth]{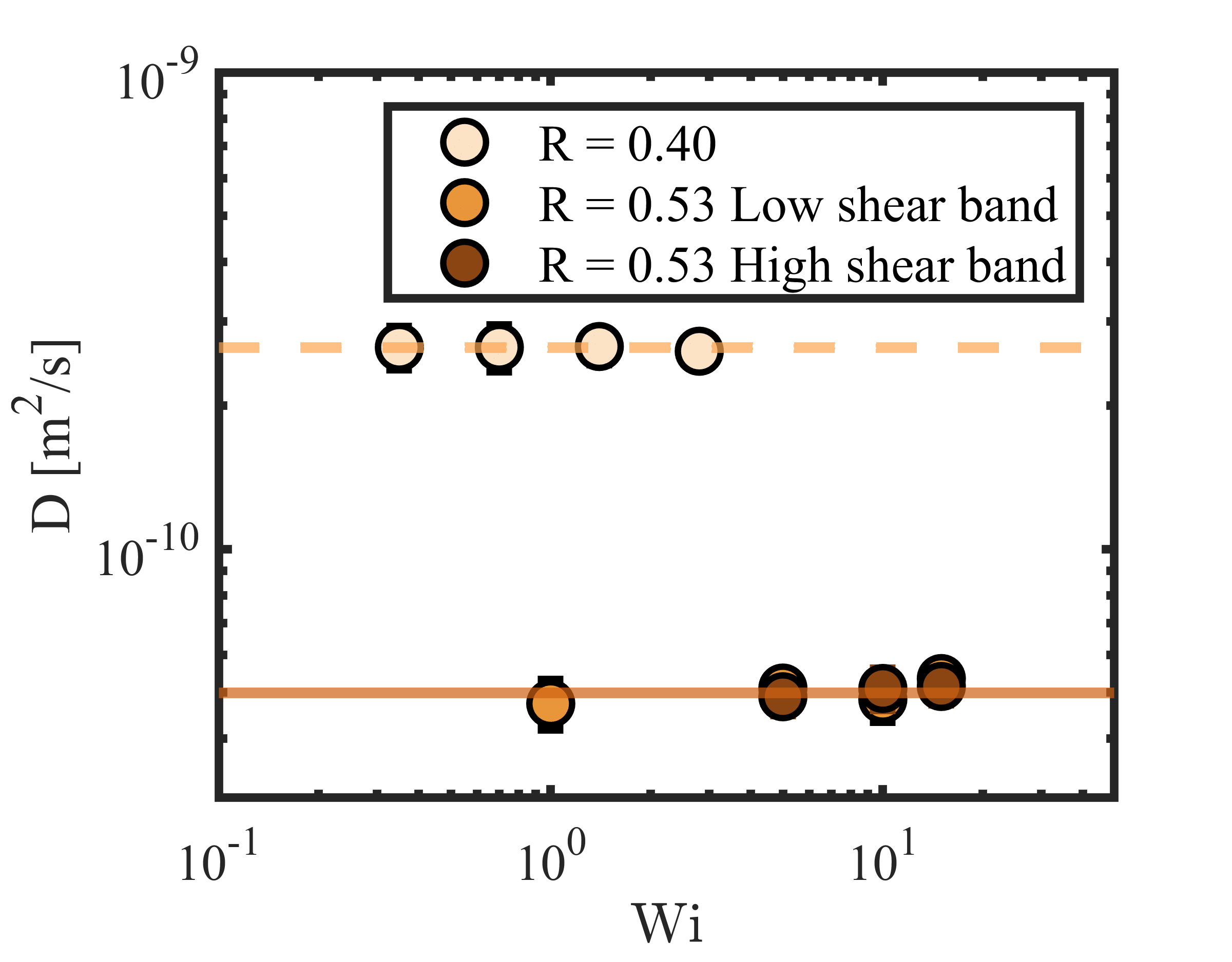}
\caption{Spatially averaged diffusion coefficient of surfactant CPyCl as a function of imposed shear rate for WLMs with $R=0.4$, and $R=0.53$. The dashed- and continuous-lines show the measured diffusivity under quiescent conditions via an independent DOSY method. }
\label{fig: averagedD}
\end{figure}

 
\HM{More remarkably, as shown in Figure~\ref{fig: dPGSE_velocity}(e,f), the surfactant self-diffusion coefficient remains spatially uniform across the gap and exhibits no measurable dependence on the imposed $Wi$ numbers for either wormlike micellar system.} This behavior persists despite the presence of pronounced shear banding in the wormlike micellar solutions with $R=0.53$ and the large variations in local shear rate that develop across the flow field. The absence of any detectable variation in surfactant self-diffusivity suggests that the local molecular mobility of the surfactant remains largely unchanged throughout the shear-banding flow. To further analyze these experimental data, we plot the gap-averaged diffusivity $D$ of CPyCl surfactants as a function of imposed $Wi$ number for both WLMs with $R=0.4$ and $R=0.53$ in Figure~(\ref{fig: averagedD}) . \HM{The numerical values of the averaged diffusion coefficients along with standard deviations are reported in Table~SI of the supplementary materials.}  Clearly, surfactant diffusivity is uniform and independent of the imposed $Wi$ numbers. For the $R=0.4$ system, the surfactant self-diffusivity under quiescent conditions was independently measured using conventional diffusion-ordered spectroscopy (DOSY) as $D\approx 2.7\times10^{-10}$ {m}$^2$/s (denoted by the dashed-line) \HM{and as shown in Fig.~S8 of the supplementary materials}. This independently measured gap-averaged diffusivity value under quiescent conditions is in close agreement with those obtained from the spatially resolved rheo-DOSY measurements under flow. In addition, Fig.~(\ref{fig: averagedD}) shows the averaged surfactant diffusivity for high and low shear bands in the shear banding WLMs with $R=0.53$. The averaged surfactant diffusivity for the high and low shear bands remains statistically indistinguishable for various imposed shear rates. Moreover, independent bulk averaged DOSY measurements produce a surfactant self-diffusivity $D \approx 4.7\times10^{-11}$ {m}$^2$/s (denoted here by the continuous line, \HM{and as shown in Fig.~S8 of the supplementary materials}), in close agreement with those values measured under flow via velocity-compensated NMR measurements. While these results are anticipated for the WLMs with $R=0.4$, where shear-induced micellar scission is unlikely, the same observation in the shear-banding WLMs is striking.\par

\begin{figure*}[ht]
\centering
\includegraphics[width=0.45\linewidth]{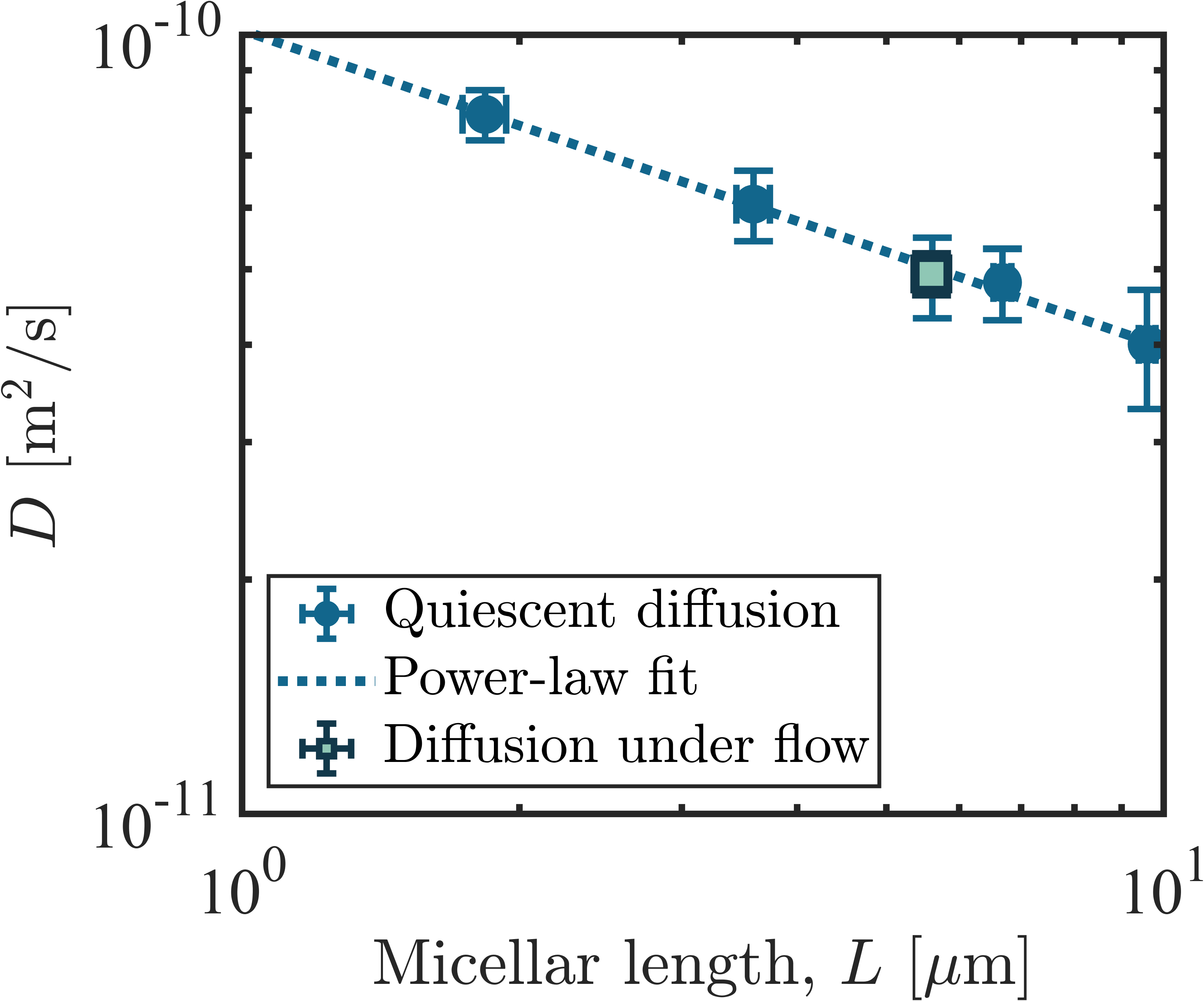}
\caption{Surfactant self-diffusion coefficient as a function of micellar length plotted for quiescent experiments, and diffusion coefficients measured for shear banding wormlike micelles with $R=0.53$. The dashed line corresponds to $D\propto L^{-0.4}.$}
\label{fig: diffvsL}
\end{figure*}

\HM{To further investigate the relationship between surfactant self-diffusion and micellar microstructure under flow, we estimated the micellar contour length and correlated it with the measured diffusion coefficients. For the strongly viscoelastic samples, for which linear viscoelastic scaling models provide reliable estimates of the mean contour length $\langle L\rangle$, equilibrium ensemble-averaged lengths were calculated using the relations introduced earlier in the manuscript. Table~SII in the supplementary materials summarizes the estimated microstructural properties of the micellar solutions at different salt-to-surfactant ratios. Figure~\ref{fig: diffvsL} presents the measured diffusion coefficients as a function of $\langle L\rangle$. Fitting the quiescent DOSY measurements yields an empirical scaling law, $D \propto \langle L\rangle^{-0.4}$. This result indicates that surfactant self-diffusivity decreases systematically with increasing micellar length. We note that other structural parameters, including mesh size and entanglement density, also vary with salt to surfactant ratio $R$. However, our estimates indicate that the changes in mesh size are comparatively modest, which suggest that variations in micellar length and entanglement density dominate the observed trends. \par 

Superimposed on this scaling are the spatially-averaged diffusion coefficient of surfactants measured for the shear-banding micellar solution under flow. Within experimental uncertainty, no statistically significant variation in $D$ is observed between flow and quiescent conditions. It is worth noting that the weak scaling exponent ($-0.4$) implies some limitations in sensitivity of diffusivity to changes in micellar length. Specifically, in flow experiments, we observe about a 10\% variations in diffusion coefficients across gap and at different imposed $Wi$ numbers. This variation in diffusion coefficient corresponds to an approximately 30\% change in the estimated mean contour length. Therefore, structural changes of 30\% are difficult to distinguish from experimental scatter. Consequently, the absence of measurable changes in $D$ does not necessarily imply an unchanged microstructure. Rather, it indicates that any flow-induced changes in the ensemble-averaged contour length are insufficiently large to produce a statistically resolvable shift in diffusivity. Therefore, our results rule out severe flow-induced micellar chain scission, while highlighting that moderate changes in micellar length may remain below the detection limit of ensemble-averaged DOSY measurements.}\par

Previous work by Cusano et al.~\cite{cusano2025exploring} reported that CPCl/Diclo wormlike micellar solutions exhibit pronounced strain hardening during step-shear flow once exceeding a critical Weissenberg number based on micellar breakage time ($Wi_b = \tau_b\dot{\gamma}$). For our shear-banding system, the micellar breakage time is estimated in Table~(\ref{Table:Rheology}). Calculating the corresponding breakage-based Weissenberg numbers yields $0.7 < Wi_b < 2.1$ across the stress plateau region. However, even at our highest imposed $Wi_b$, our experiments reveal no significant strain hardening. Although our concentrations account for the density of D$_2$O, differences in solvent composition and density should be considered when comparing these results with with literature data from H$_2$O-based systems, as structural and rheological transitions may occur at different nominal concentrations. Our results are consistent with predictions from tube-based models for shear-banding wormlike micellar solutions~\cite{spenley1993nonlinear,peterson2021predictions}. In particular, Peterson and Leal developed a constitutive model based on reptation-reaction dynamics and micellar scission/recombination kinetics, and showed that a stress plateau and shear banding can emerge solely from equilibrium micellar breaking and reformation. The model did not require the inclusion of additional flow-induced micellar breakage to reproduce these features. This prediction is consistent with our observations. Hence although the present wormlike micellar system of this study may lie in the fast-breaking regime, the diffusion measurements do not provide evidence for significant flow-induced micellar breakage under shear-banding conditions.
To shed more light on these experimental observations, the velocity-compensated signal attenuation measured under flow is used to determine the mean-squared displacement of the molecules (MSD), $Z^2(t)$:

\begin{equation} 
 \frac{S(q)}{S_0} = \exp\left[-\frac{1}{2}Z^2(t)q^2\right],
 \label{MSD}
\end{equation}

\begin{figure}[ht]
\centering
\includegraphics[width=1\linewidth]{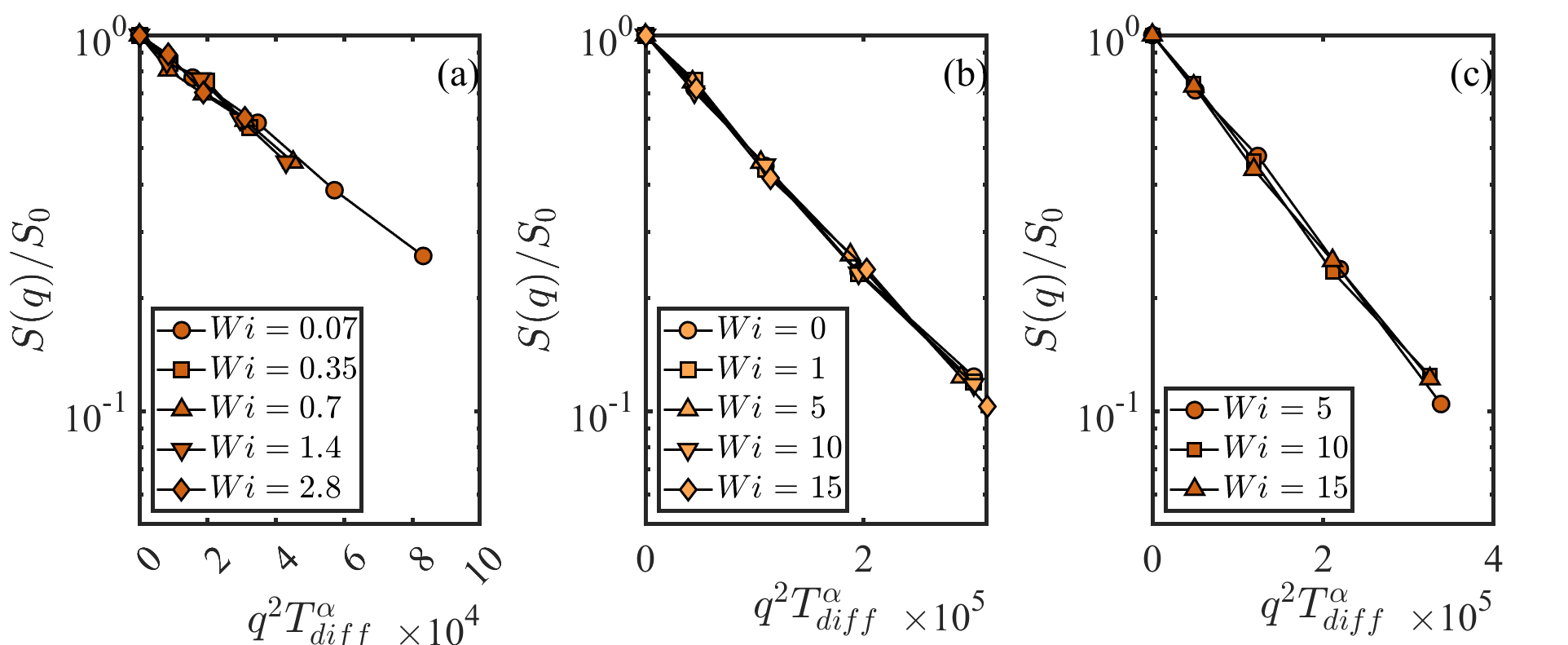}
\caption{Normalized NMR signal attenuation of surfactant peaks as a function of diffusion weighting for various imposed shear rates. The left subplot corresponds to WLMs with $R = 0.4$\HM{ with $\alpha =0.71\pm0.04$ }, and the middle and right side panels correspond to low shear band (middle panel), and high shear band (right panel) of the shear banding WLMs \HM{with $\alpha = 0.5\pm0.01$ for low shear band and $\alpha = 0.49\pm0.03$ for high shear band}. These data are superimposed by a common $\alpha$ parameter denoting the level of restriction in their diffusivity.}
\label{fig: alpha_superimposed}
\end{figure}

Previous studies have demonstrated that the scaling exponent $\alpha$, obtained from the dependence of the mean-squared displacement on diffusion time, is highly sensitive to micellar morphology in wormlike micellar solutions~\cite{scigliani2023probing,holder2021nuclear,angelico1998surfactant}. To determine $\alpha$ under flow conditions, the normalized signal attenuation was plotted as a function of $q^2T_{\mathrm{diff}}^{\alpha}$, and $\alpha$ was adjusted until all attenuation data collapsed onto a single master curve. Figure~\ref{fig: alpha_superimposed}(a) shows the normalized signal attenuation for the $R=0.4$ system over a range of imposed shear rates. An excellent collapse of the data is obtained with an average exponent of $\alpha = 0.71 \pm 0.04$. This value is consistent with a heterogeneous microstructure comprising a mixture of a small fraction of spherical micelles, short wormlike micelles, and longer entangled micelle chains, in agreement with previous diffusion-NMR studies of surfactant systems in this compositional regime. Figures~\ref{fig: alpha_superimposed}(b,c) present the corresponding analysis for the $R=0.53$ shear-banding system in the low-shear (part a) and high-shear bands (part b), respectively. Remarkably, the normalized signal attenuation from all imposed shear rates collapses onto a common master curve with $\alpha \approx 0.5$ in both bands. This value is characteristic of sub-diffusive curvilinear diffusion of surfactants in an entangled wormlike micelle network~\cite{angelico1998surfactant}. More importantly, no measurable difference in $\alpha$ is observed between the low-shear and high-shear bands, indicating that the local micellar morphology remains essentially unchanged across the shear-banding interface. \HM{Additionally, these values of MSD exponents are consistent with those measured under quiescent conditions using the DOSY experiments (c.f. Fig.~S9, in the supplementary materials).}\par  

Finally, Fig.~\ref{fig: alpha} shows the variation of the MSD power-law index $\alpha$ measured as a function of imposed shear rate for both WLMs. For the shear banding system, and for all imposed $Wi$ numbers, $\alpha = 0.5 \pm 0.01$ for both low and high shear bands. These results indicate that surfactant self-diffusion in shear-banding WLMs follows curvilinear diffusion along the micelle chains under flow conditions with $\alpha \approx0.5$~\cite{angelico1998surfactant}, and the mechanisms of self-diffusion. Together with the self-diffusion measurements, these results suggest that, although shear banding produces substantial heterogeneity in the velocity and shear-rate fields, it does not lead to measurable changes in either the translational diffusivity or the microstructural dynamics of the surfactant molecules, as quantified by the scaling exponent $\alpha$. The invariance of both the self-diffusion coefficient and $\alpha$ across the TC gap suggests that the micellar microstructure remains similar in the low- and high-shear bands. This observation further suggests that a significant flow-induced micellar chain scission is negligible in this shear-banding system for imposed $Wi$ numbers within the stress-plateau regime, despite the large differences in local shear rate between the coexisting bands.\par 

\HM{Beyond flow-induced micellar scission, other mechanisms have been proposed to explain shear banding in semidilute wormlike micellar solutions. Some models predict flow–concentration coupling, in which the high- and low-shear bands have different micellar concentrations~\cite{schmitt1995shear,fielding2003flow}. Other studies attribute shear banding to flow-induced micellar disentanglement or flow-induced micellar alignment analogous to mechanisms proposed for entangled polymer solutions~\cite{peterson2021predictions,spenley1993nonlinear,Lopez2014,Mc24}. For fast-breaking wormlike micelles at fixed temperature, concentration and at equilibrium conditions, the degree of micellar entanglement is closely related to the micellar contour length as noted above in the manuscript. Thus, a substantial reduction in entanglement would generally be expected to accompany a decrease in micellar length, unless flow promotes disentanglement without significant micellar scission. Our results show no evidence of substantial flow-induced changes in micellar length. They therefore do not exclude either flow-induced disentanglement without significant scission or flow–concentration coupling as possible mechanisms that give rise to shear banding. Distinguishing between these mechanisms will require direct measurements of local micellar concentration and entanglement within the individual shear bands.}\par 

\begin{figure}[ht]
\centering
\includegraphics[width=0.5\linewidth]{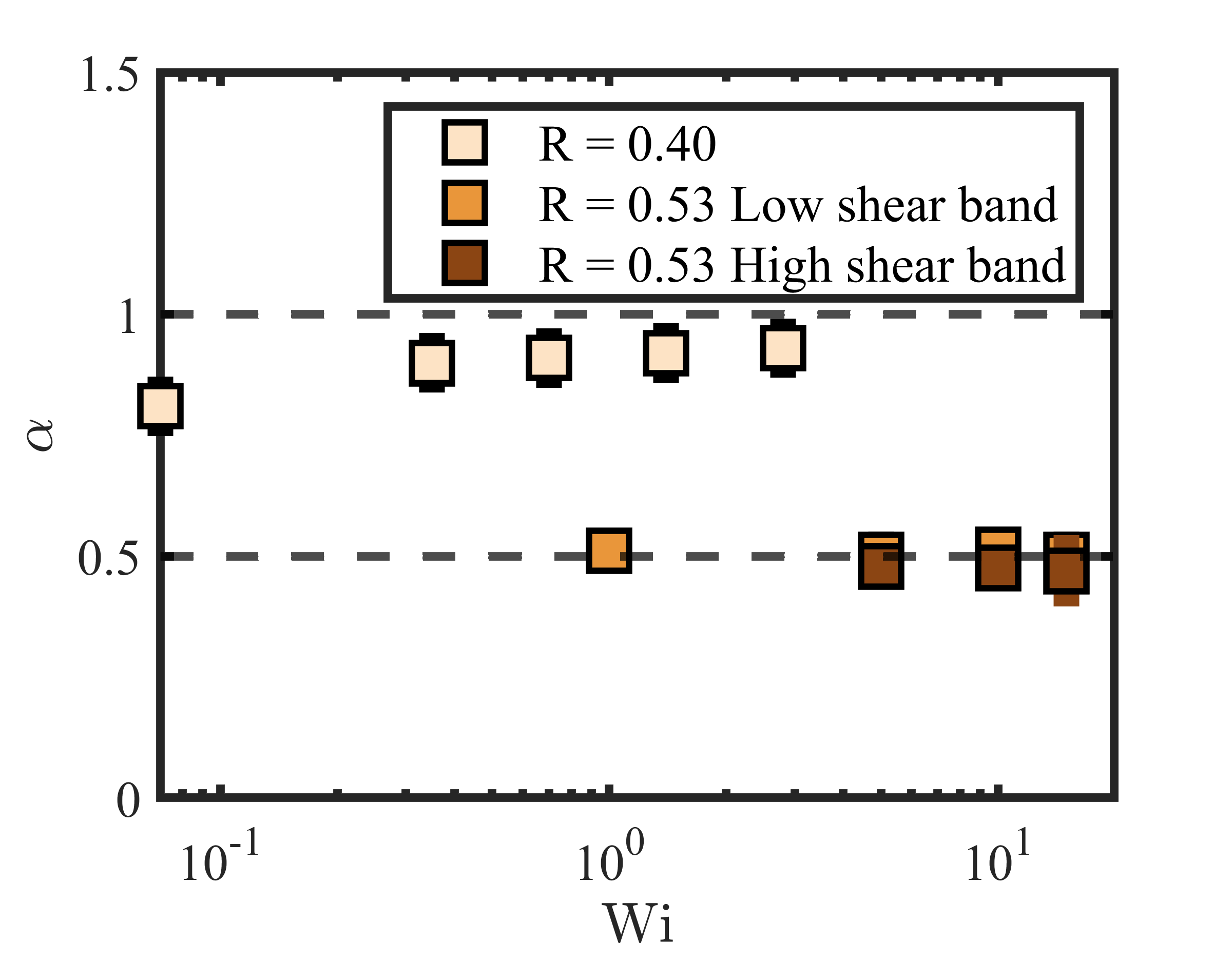}
\caption{Mean squared displacement index $\alpha$ as a function of imposed shear rates for WLMs with salt-to-surfactant ratio of $R=0.4$, and $R=0.53$. }
\label{fig: alpha}
\end{figure}

\section{Summary}
In summary, we combined rheo-optical imaging and spatially resolved rheo-DOSY-NMR measurements to directly investigate the possible role of flow-induced micellar chain scission in both a weakly viscoelastic shear-thinning wormlike micellar solution and a strongly shear-banding system based on CPyCl/NaSal/D$_2$O. Rheo-optical observations and velocity measurements confirmed the distinct flow responses of the two fluids, ranging from homogeneous shear flow in the non-shear-banding system to the coexistence of low- and high-shear bands within the stress-plateau regime of the shear-banding fluid. By employing velocity-compensated diffusion measurements, we were able to isolate molecular diffusion from advective transport and quantify the local surfactant self-diffusivity across the Taylor--Couette gap under flow. Despite the pronounced spatial heterogeneity in velocity and shear-rate fields associated with the shear banding system, the measured surfactant self-diffusivity remained uniform across the gap and showed no systematic dependence on the imposed shear rate. Furthermore, analysis of the mean-squared-displacement scaling revealed the diffusion exponent remained essentially unchanged throughout the flow field, including within the high- and low-shear bands. These results suggest that the underlying micellar microstructure in this shear banding system remains largely preserved, even in regions experiencing substantially different local shear rates. This finding is particularly significant because flow-induced breakage has frequently been invoked as a potential mechanism underlying shear band formation and stress-plateau behavior in wormlike micellar solutions.\par 

The conclusions of the present study are consistent with tube-based constitutive theories of wormlike micelles~\cite{spenley1993nonlinear,peterson2021predictions}, as well as more recent theoretical descriptions of shear banding~\cite{salipante2024two}, which do not require extensive flow-induced micellar chain scission to reproduce the shear-banding flow phenomenon. More broadly, these results provide direct experimental constraints for future theoretical and numerical models of shear banding and highlight the importance of mechanisms other than flow-induced micellar breakage in governing the emergence and dynamics of shear bands. 

\section{Acknowledgment}
{We acknowledge financial support from NSF through the award CBET CAREER 1942150. NMR experiments were performed at the US National High Magnetic Field Laboratory (NHMFL), which is supported by the State of Florida and the National Science Foundation Cooperative Agreement No. DMR-1644779.}

\bibliography{bib} 
\bibliographystyle{rsc} 
\newpage

\section*{\centering Appendix: Supplementary Information}
\setcounter{figure}{0}
\renewcommand{\thefigure}{S\arabic{figure}}
\renewcommand{\theHfigure}{S\arabic{figure}}

\setcounter{table}{0}
\renewcommand{\thetable}{S\Roman{table}}
\renewcommand{\theHtable}{S\Roman{table}}

\begin{figure}[ht]
\centering
\includegraphics[width=0.8\linewidth]{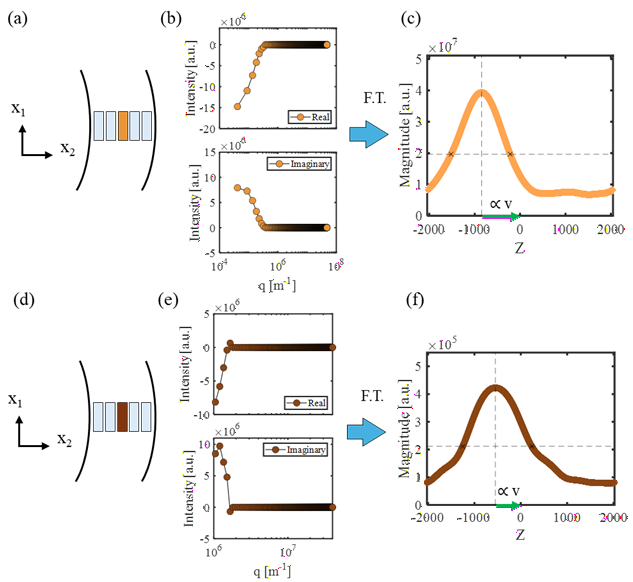}
\caption{A step-by-step method used for the acquisition of velocity images. Top row: Calibrations with Newtonian fluid for $\dot\gamma = 20$ [1/s] at $x_2/d = 0.5$, and bottom row results for shear banding wormlike micelles at $Wi$ = 15 at $x_2/d = 0.5$.  \HM{(a,d) Selected voxels within the TC cell gap. (b,e) Real and imaginary components of the q-space Free Induction Decay (FID). (d,f) Dynamic displacement profile (DDP) where the horizontal displacement is indicated in green, and it is proportional to the velocity of the voxel as noted under Rheo-DOSY NMR section of the paper. In this method, the Displacement Density Profile (DDP) is calculated by applying a Fourier transform to the complex MRI signal measured across varying diffusion-encoding gradient strengths ($q$-space). For each target voxel, the real and imaginary raw DICOM components are combined into a complex signal $S(q)$. This $q$-space signal is Fourier-transformed into displacement space, and its magnitude is taken to generate the DDP. The mean shift of the resulting profile is then used to determine the local bulk velocity within each voxel. }}
\label{fig: CallaghanMethod}
\end{figure}

\begin{figure}[ht]
\centering
\includegraphics[width=0.45\linewidth]{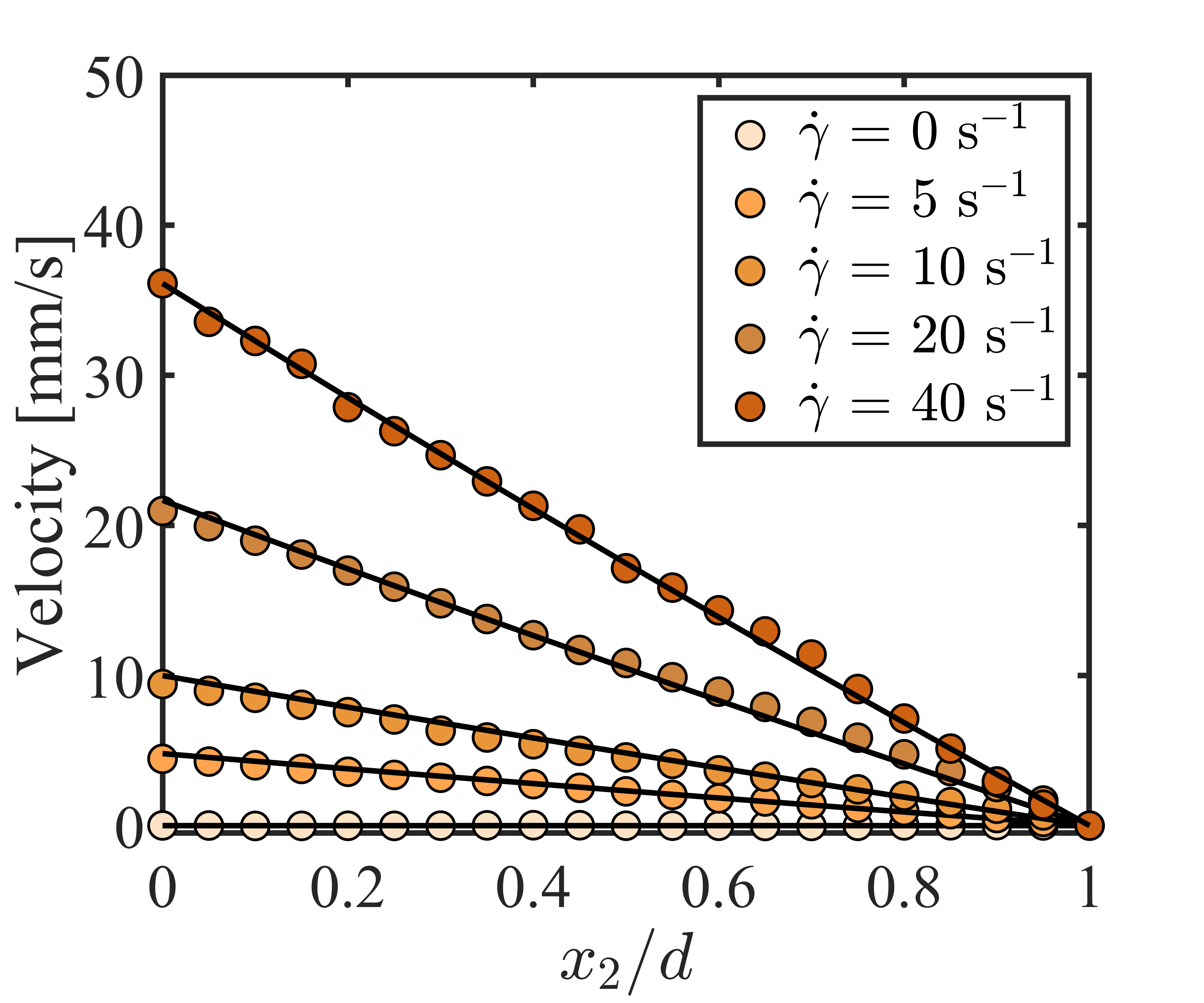}
\caption{PGSE calibration with Newtonian Fluid. 1D temporally averaged velocity profiles as a function of radial position for multiple imposed shear rates. The lines correspond to predictions based on the Navier-Stokes equation.}
\label{fig: PGSE_Newtonian}
\end{figure}

\begin{figure}[ht]
\centering
\includegraphics[width=0.75\linewidth]{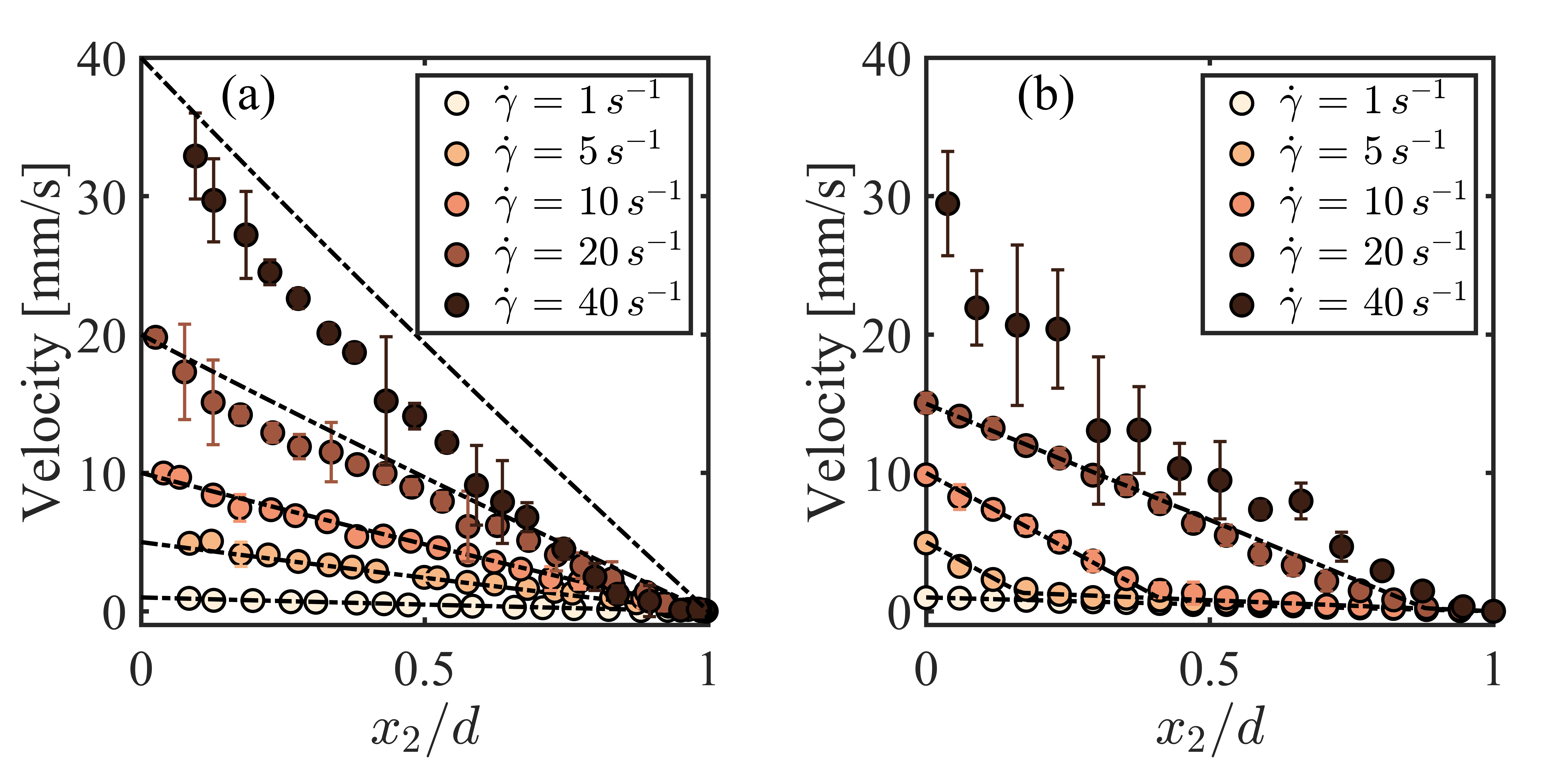}
\includegraphics[width=0.75\linewidth]{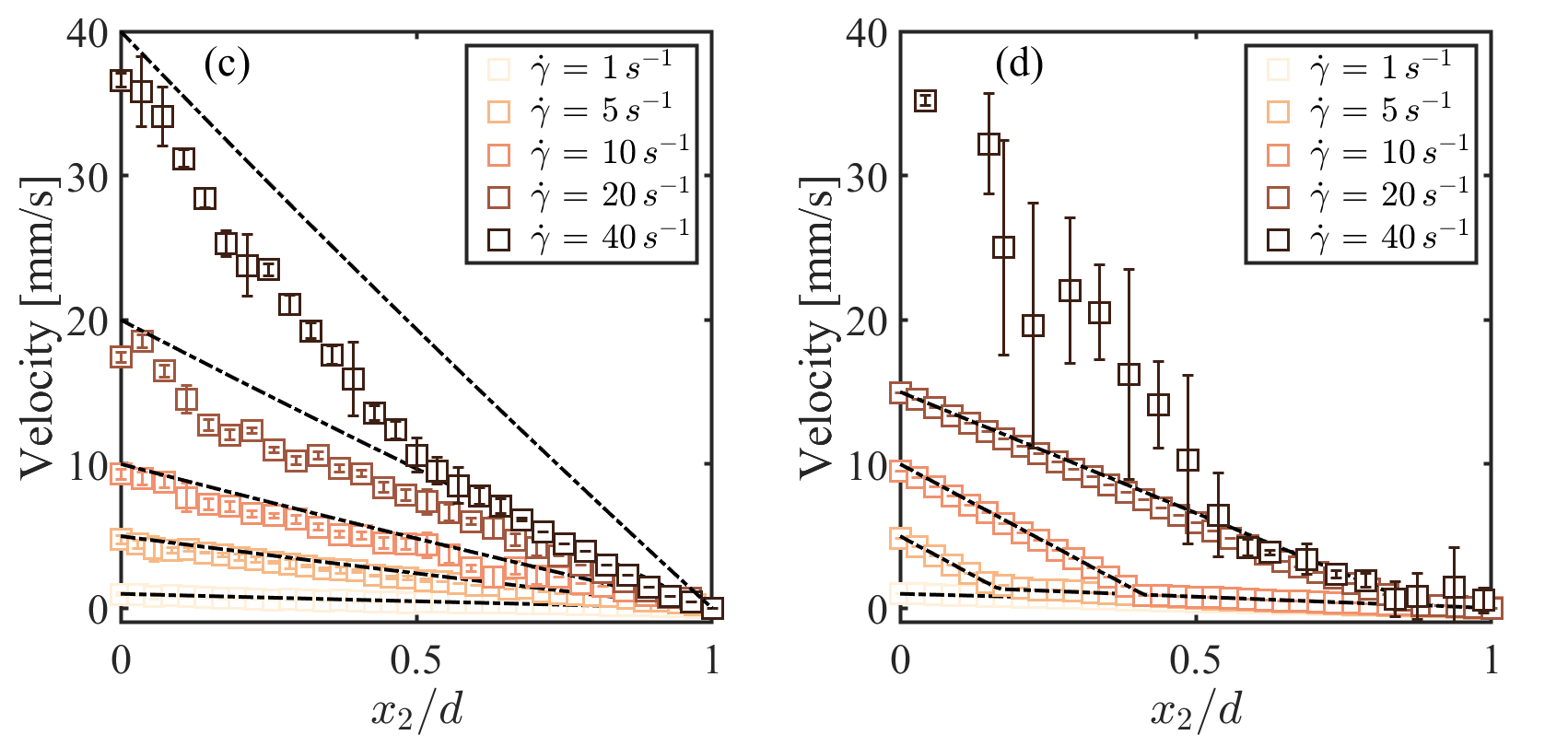}

\caption{1D time-averaged velocity profiles
for weakly viscoelastic WLMs (a,c) and shear-banding WLMs (b,d) measured by rheo-optics (c,d)
and PGSE-NMR (a,b). Dashed and Black lines show predictions from the Newtonian
behavior and lever rule, respectively}
\label{fig: separate}
\end{figure}

\begin{figure}[ht]
\centering
\includegraphics[width=0.8\linewidth]{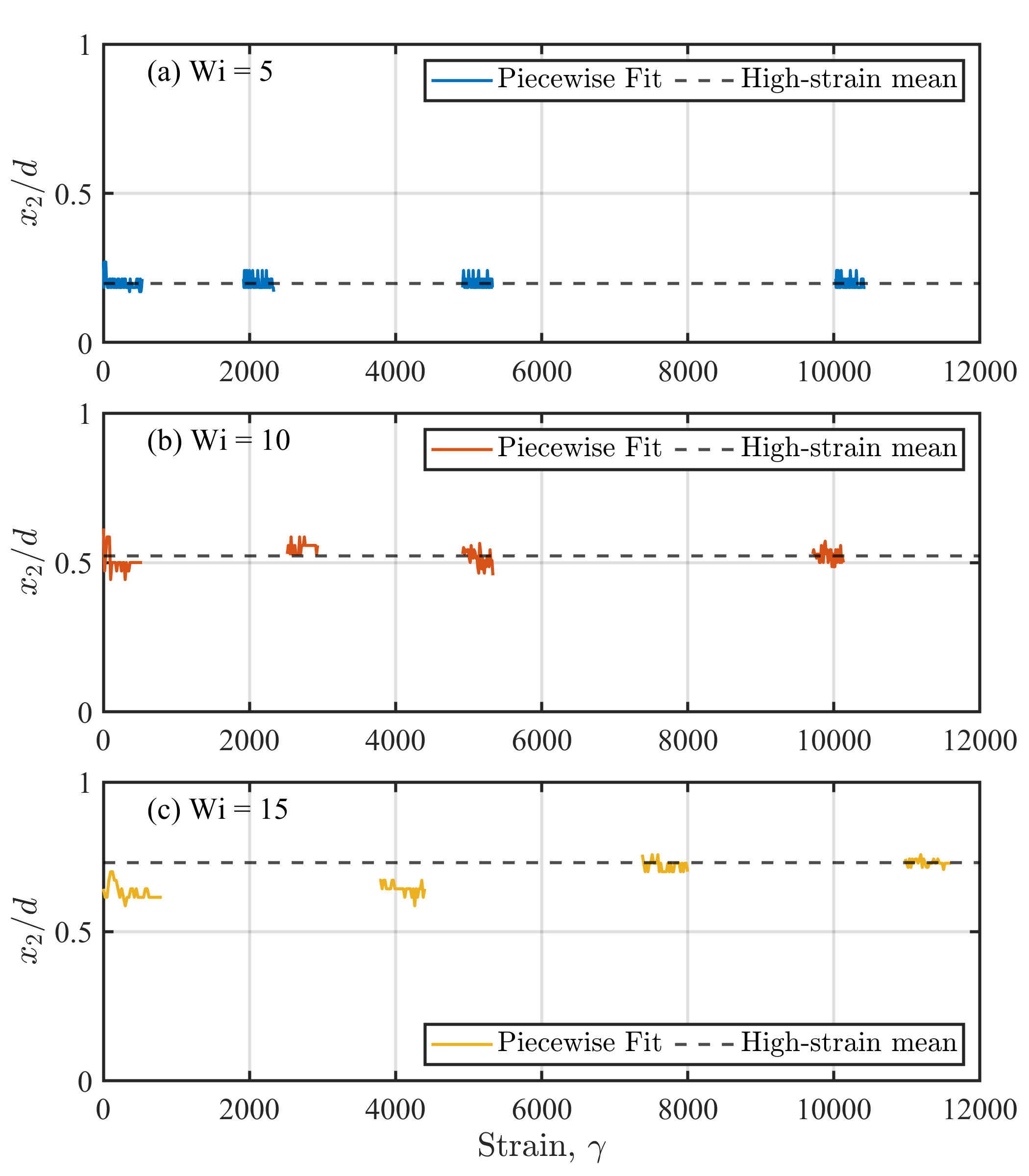}
\caption{Evolution of the normalized interface position ($x_2/d$) with imposed strain ($\gamma$) for WLMs with $R=0.53$ at (a) $Wi=5$, (b) $Wi=10$, and (c) $Wi=15$. Solid lines represent the piecewise-linear fits, while dashed horizontal lines indicate the mean interface position at high strain.}
\label{fig: Interface_NMR}
\end{figure}

\begin{figure}[ht]
\centering
\includegraphics[width=0.75\linewidth]{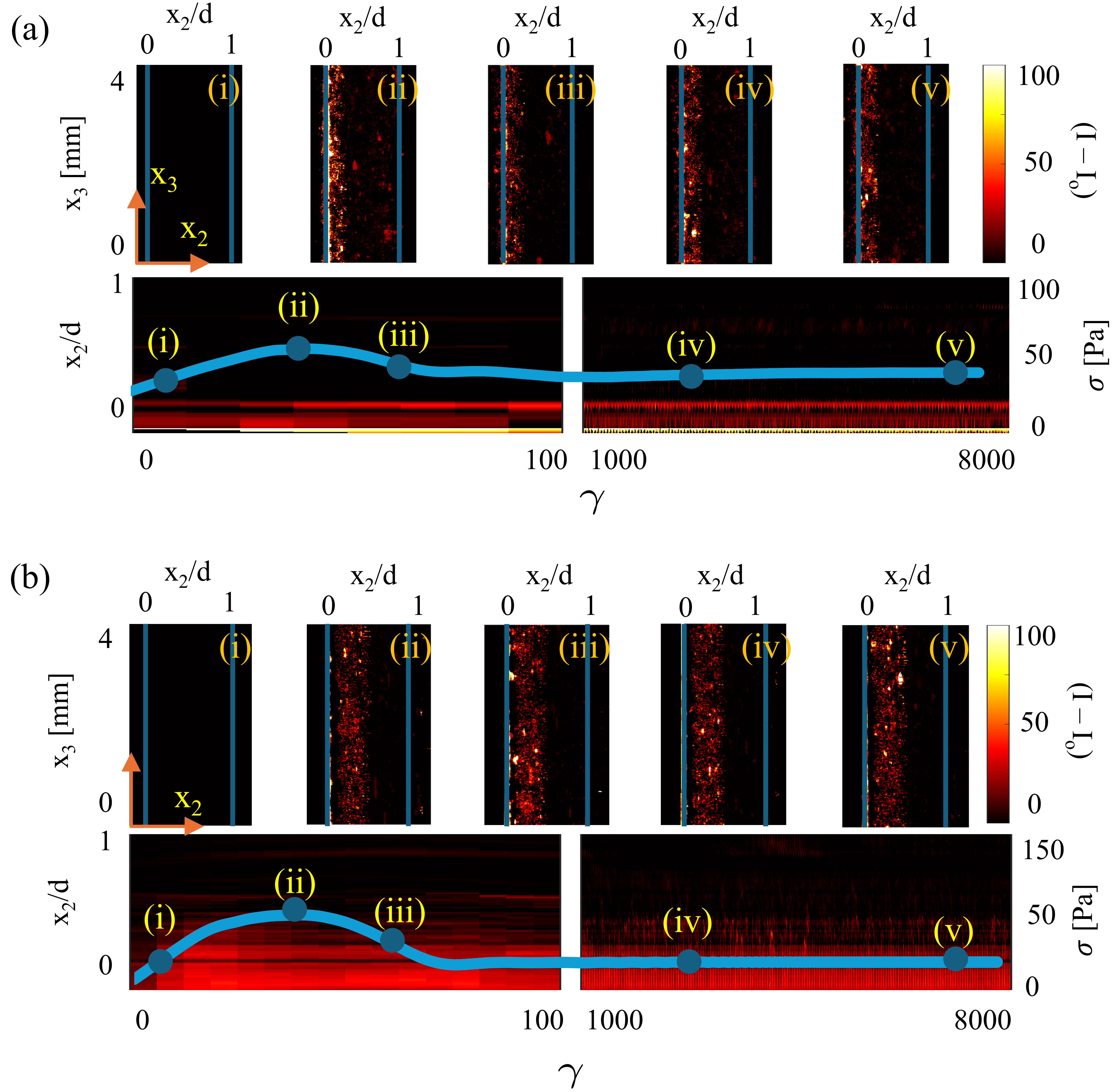}
\caption[Turbidity in the $x_2-x_3$ plane ]{Turbidity measured in the $x_2-x_3$ plane as a function of strain $\gamma$, where $\gamma$ = $\dot\gamma t$ is used as a dimensionless representation of time. Top row: Snapshot frames. Bottom row: shear stress evolution overlaid on spatiotemporal display of turbidity. For imposed $Wi$ of (a) 5, and (b) 10}
\label{fig: Turbidity_RZ}
\end{figure}

\begin{figure}[ht]
\centering

\includegraphics[width=0.8\linewidth]{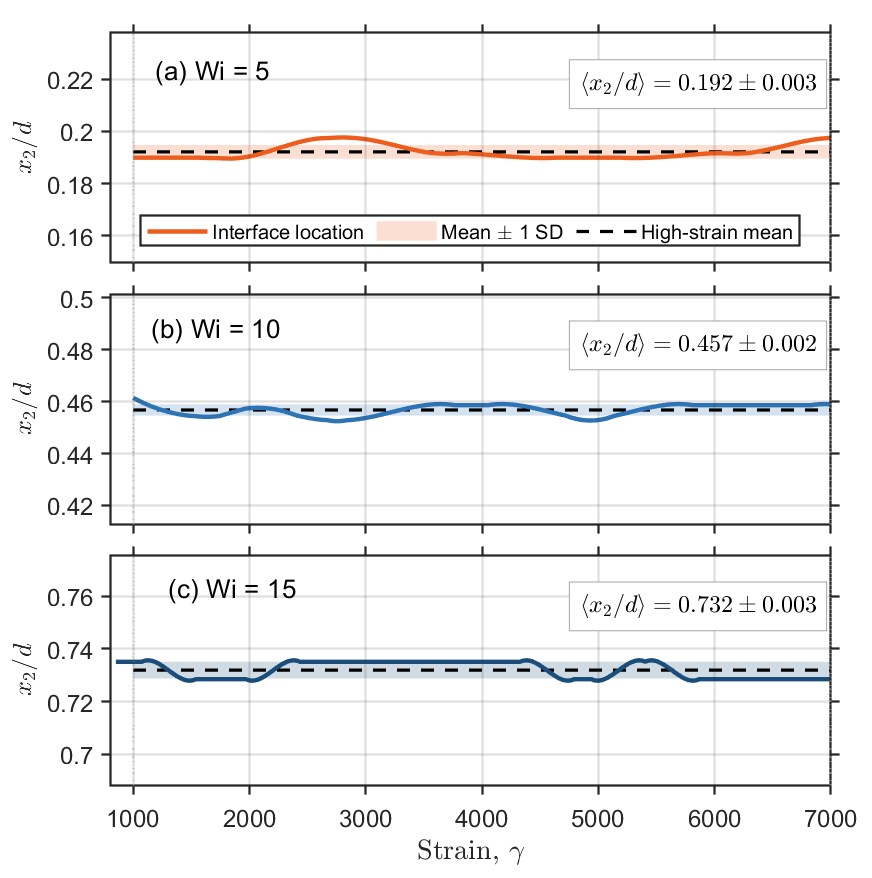}
\caption{Interface between high and low shear bands as a function of strain for the shear banding wormlike micellar solutions at different imposed $Wi$ numbers.}
\label{fig: Interface}
\end{figure}

\begin{figure}[ht]
\centering
\includegraphics[width=0.4\linewidth]{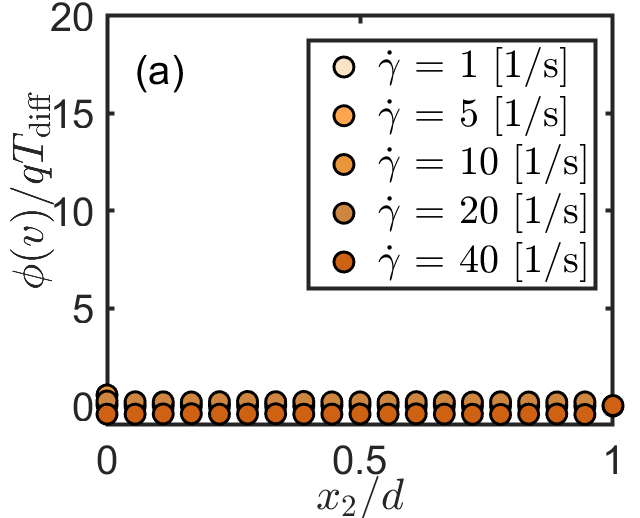}
\caption{Velocity-compensated signal attenuation arising from advective flow ($\phi (v)= qT_\mathrm{diff}v$) across the TC gap at various imposed shear rates for WLMs with $R=0.4$.}
\label{fig: dPGSE_Newtonian}
\end{figure}

\begin{table}[ht] \centering \caption{Mean diffusion coefficients and standard deviations at different shear conditions.} \label{tab:adc_shear_rate} \footnotesize \setlength{\tabcolsep}{6pt} \renewcommand{\arraystretch}{1.15} \begin{tabular}{cccccc} \toprule \multicolumn{2}{c}{$R=0.40$} & \multicolumn{2}{c}{$R=0.53$: Low-shear band} & \multicolumn{2}{c}{$R=0.53$: High-shear band} \\ \cmidrule(lr){1-2} \cmidrule(lr){3-4} \cmidrule(lr){5-6} $Wi$ & $D$ [m$^2$/s] & $Wi$ & $D$ [m$^2$/s] & $Wi$ & $D$ [m$^2$/s] \\ \midrule 0.07 & $(2.67 \pm 0.32)\times10^{-10}$ & 0.00 & $(4.99 \pm 0.36)\times10^{-11}$ & 0.00 & -- \\ 0.35 & $(2.65 \pm 0.24)\times10^{-10}$ & 1.00 & $(4.74 \pm 0.51)\times10^{-11}$ & 1.00 & -- \\ 0.70 & $(2.65 \pm 0.27)\times10^{-10}$ & 5.00 & $(5.11 \pm 0.32)\times10^{-11}$ & 5.00 & $(4.90 \pm 0.36)\times10^{-11}$ \\ 2.80 & $(2.66 \pm 0.19)\times10^{-10}$ & 10.00 & $(4.84 \pm 0.45)\times10^{-11}$ & 10.00 & $(5.10 \pm 0.45)\times10^{-11}$ \\ 2.80 & $(2.60 \pm 0.16)\times10^{-10}$ & 15.00 & $(5.35 \pm 0.30)\times10^{-11}$ & 15.00 & $(5.15 \pm 0.38)\times10^{-11}$ \\ \bottomrule \end{tabular} \end{table}

\begin{table}[ht] \centering \caption{Micellar properties at different NaSal-to-CPyCl ratios. These properties are estimated using the equations noted in the main manuscript.} \label{tab:micellar_properties} \begin{tabular}{ S[table-format=1.2] S[table-format=2.2] S[table-format=1.2] S[scientific-notation=true, table-format=1.2e-1] S[table-format=3.2] S[table-format=2.2] S[table-format=2.2] } \toprule {$R$} & {$G_o$ [Pa]} & {$G''_{\min}$ [Pa]} & {$l_e$ [m]} & {$Z$} & {$\langle L\rangle$ [$\mu$m]} & {$\xi$ [nm]} \\ \midrule 0.45 & 24.00 & 6.71 & 9.11e-8 & 19.17 & 1.75 & 56.48 \\ 0.50 & 33.70 & 4.50 & 7.54e-8 & 47.21 & 3.56 & 50.44 \\ 0.53 & 35.00 & 3.50 & 7.79e-8 & 67.18 & 5.23 & 49.81 \\ 0.55 & 36.45 & 2.50 & 6.87e-8 & 106.39 & 7.31 & 49.14 \\ 0.60 & 45.00 & 2.40 & 7.18e-8 & 144.59 & 10.38 & 45.81 \\  \bottomrule \end{tabular} \end{table}

\begin{figure}[ht]
\centering

\includegraphics[width=0.99\linewidth]{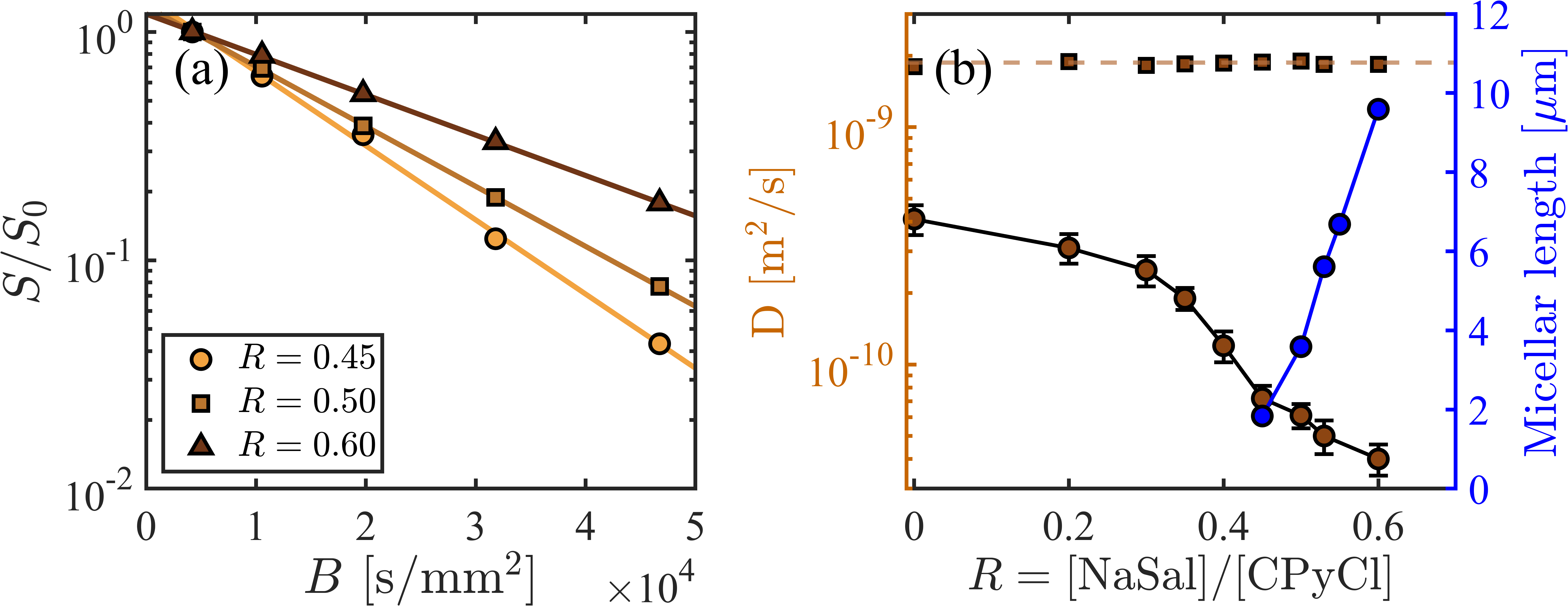}
\caption{(a) Normalized NMR signal attenuation as a function of diffusion weighting for micellar solutions at different salt-to-surfactant concentration ratios, $R$. Solid lines represent monoexponential fits. The lines show the fit to a monoexponential type decay. (b) Surfactant self-diffusion coefficient and average micellar length as functions of $R$. The left and right y-axes show the diffusion coefficient and micellar length, respectively. Circles represent surfactant self-diffusion coefficients, and squares represent residual water diffusion coefficients. }
\label{fig: dPGSE_Newtonian_SI}
\end{figure}

\begin{figure}[ht]
\centering
\includegraphics[width=0.60\linewidth]{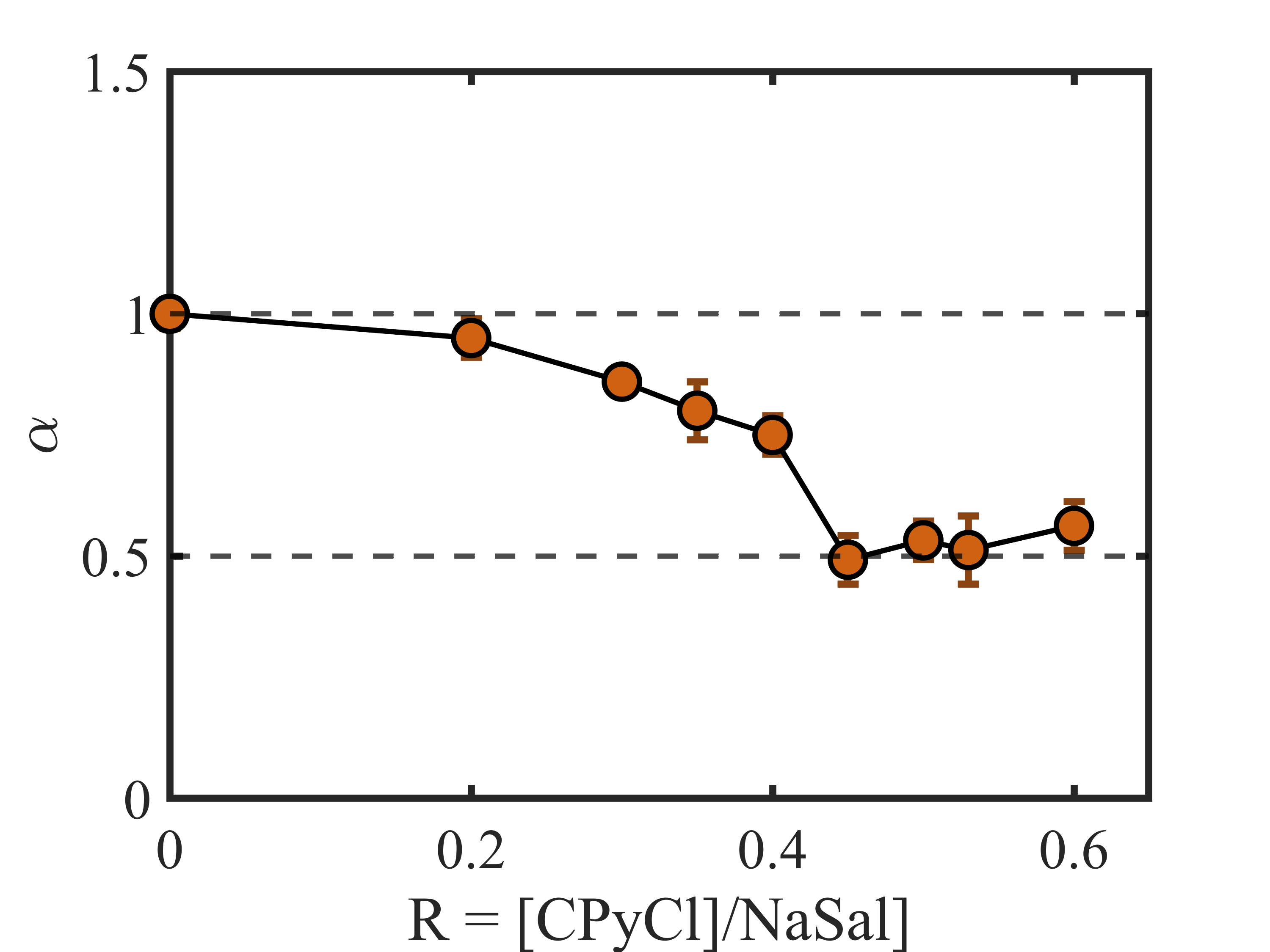}
\caption{MSD exponent $\alpha$ measured for quiescent wormlike micellar solutions as a function of salt to surfactant concentration ratio.}
\label{fig: Alpha_R}
\end{figure}

\end{document}